\documentclass[12pt]{article}
\usepackage{epsf}
\def\lsim{\mathrel{\rlap {\raise.5ex\hbox{$ < $}}
{\lower.5ex\hbox{$\sim$}}}}
\def\gsim{\mathrel{\rlap {\raise.5ex\hbox{$ > $}}
{\lower.5ex\hbox{$\sim$}}}}

\def\sqr#1#2{{\vcenter{\vbox{\hrule height.#2pt
        \hbox{\vrule width.#2pt height#1pt \kern#1pt
           \vrule width.#2pt}
        \hrule height.#2pt}}}}

\def\lsim{{\displaystyle
{{\raise-8pt\hbox{$ <$}}
\atop{\raise5pt\hbox{$\sim$}}}}}
\def\gsim{{\displaystyle
{{\raise-8pt\hbox{$ >$}}
\atop{\raise5pt\hbox{$\sim$}}}}}
\def\slsim{{\displaystyle
{{\raise-8pt\hbox{$\scriptstyle <$}}
\atop{\raise5pt\hbox{$\scriptstyle \sim$}}}}}
\def\sgsim{{\displaystyle
{{\raise-8pt\hbox{$\scriptstyle  >$}}
\atop{\raise5pt\hbox{$\scriptstyle \sim$}}}}}
\def\Tr{\,{\rm Tr}\, }
\def\Im{\,{\rm Im}\, }
\def\Re{\,{\rm Re}\, }
\def\bJ{\overline{J}}
\def\bj{\overline{j}}
\def\bI{\overline{I}}
\def\bP{\overline{P}}

\def\bE{\overline{E}}
\def\bF{\overline{F}}
\def\bT{\overline{T}}
\def\bU{\overline{U}}
\def\bOmega{\overline{\Omega}}
\def\s{\sigma}
\def\a{\alpha}
\def\g{\gamma}
\def\thefootnote{\fnsymbol{footnote}}
\def\be{\begin{equation}}
\def\ee{\end{equation}}
\def\ba{\begin{eqnarray}}
\def\ea{\end{eqnarray}}
\def\bs{\begin{subequations}}
\def\es{\end{subequations}}

\def\tb{\bar{\theta}}

\def\l{\lambda}

\def\rr{{\rm regular}}

\def\pp{\bar\p}

\def\qb{{\bar q}}

\def\t{\tau}
\def\tb{\bar\tau}

\def\RR{{\rm I\!R}}

\def\gs{g_{\rm string}}

\def\pp{\partial}
\def\ap{\alpha'}
\def\za{{Z}_{F}}

\def\f{{\cal F}}
\def\rr{{\cal R}}
\def\t{\tau}
\def\im{\, {\rm Im}\, \tau}

\def\npi#1#2#3{{\bf{B#1}} (#2) #3}
\def\np#1#2#3{Nucl. Phys. {\bf{B#1}} (#2) #3}
\def\pl#1#2#3{Phys. Lett. {\bf{#1B}} (#2) #3}
\def\prl#1#2#3{Phys. Rev. Lett. {\bf{#1}} (#2) #3}
\def\pr#1#2#3{Phys. Rev. {\bf{D#1}} (#2) #3}
\def\nl{\hfil\break}
\def\thebibliography#1{%
\vskip 0.5cm \centerline{\bf References}
\list{%
[\arabic{enumi}]}{\settowidth\labelwidth{[#1]}
\leftmargin\labelwidth
\advance\leftmargin\labelsep
\usecounter{enumi}}
\def\newblock{\hskip .11em plus .33em minus .07em}
\sloppy\clubpenalty4000\widowpenalty4000
\sfcode`\.=1000\relax}

\begin{document}
\renewcommand{\theequation}{\arabic{section}.\arabic{equation}}
\begin{titlepage}
\begin{flushright}
CERN-TH/96-90 \\
SISSA/87/96/EP \\
LPTENS/96/26 \\
hep-th/9608034 \\
\end{flushright}
\begin{centering}
\vspace{.3in}
{\bf UNIVERSALITY PROPERTIES OF $N=2$ AND $N=1$ HETEROTIC THRESHOLD
CORRECTIONS}\\
\vspace{1. cm}
{E. KIRITSIS$^{\ 1}$, C. KOUNNAS$^{\ 1,\, \ast}$,
P.M. PETROPOULOS$^{\ 1,\, \ast}$ \\
\medskip
and \\
\medskip
J. RIZOS$^{\ 1,\, 2,\, \diamond}$}\\
\vskip 1cm
{$^1 $\it Theory Division, CERN}\\
{\it 1211 Geneva 23, Switzerland}\\
\medskip
{\it and}\\
\medskip
{$^2 $\it International School for Advanced Studies, SISSA}\\
{\it Via Beirut 2-4, 34013 Trieste, Italy}\\
\vspace{1.0cm}
{\bf Abstract}\\
\end{centering}
\vspace{.1in}
In the framework of heterotic compactifications,
we consider the one-loop corrections to the gauge couplings,
which were shown  to be free of any  infra-red ambiguity.
For a class of $N=2$ models, namely those that are obtained
by toroidal compactification to four dimensions of generic
six-dimensional $N=1$ ground states, we give an explicit formula for
the
gauge-group independent thresholds, and show that these are equal
within this class, as a consequence
of an anomaly-cancellation constraint in six dimensions.
We further use these results to
compute the ($N=2$)-sector  contributions to the thresholds
of $N=1$ orbifolds.
We then consider the full contribution of $N=1$ sectors to the gauge
couplings, which generically are expected to modify the unification
picture.
We compute such corrections in several models.
We finally comment on the effect of such contributions
to the issue of string unification.
\vspace{.5cm}
\begin{flushleft}
CERN-TH/96-90 \\
SISSA/87/96/EP \\
LPTENS/96/26 \\
July 1996 \\
\end{flushleft}
\hrule width 6.7cm \vskip.1mm{\small \small \small
$^\ast$\ On leave from {\it Centre National de
la Recherche Scientifique,} France.\\
$^\diamond$\ On leave from {\it Division of Theoretical Physics,
Physics Department,}\\
$^{\ }$\ {\it University of Ioannina,} Greece.}
\end{titlepage}
\newpage
\setcounter{footnote}{0}
\renewcommand{\thefootnote}{\arabic{footnote}}

\setcounter{section}{1}
\section*{\normalsize{\bf 1. Introduction}}

\noindent
In the past several years, there has been significant progress in
trying to compare
low-energy predictions of string theory with data
\cite{gin}--\cite{BFO}.
String theory gives us the possibility of unifying gauge, Yukawa and
gravitational interactions.
The presence of supersymmetry is usually required in order to deal
with hierarchy problems, although in the context of supergravity
this is not automatic due to the presence of gravity (this can be
verified
in the framework of strings, where gravitational interactions are
properly taken into account \cite{fkz}).
The standard folklore demands $N=1$ supersymmetry in order for the
theory to possess chiral fermions.
Extended-supersymmetry ground states could also be considered
provided the supersymmetry is spontaneously broken to
$N=1$  \cite{deco, kka}. Indeed, there are indications that such
theories
become chiral at some special points of the string moduli space.

The quantities that are most easily comparable to experimental data
are effective gauge couplings of the observable sector, as well as
Yukawa couplings.
It is well known that the low-energy world is not supersymmetric.
Thus ($N=1$) supersymmetry has to be broken spontaneously at some
scale of
the order of 1 TeV (for hierarchy reasons).
Although there are ways to break supersymmetry in string theory
\cite{gau,ss,bach}, it is fair to say that none so far has yielded a
phenomenologically acceptable model.
The issue of supersymmetry breaking is therefore an open problem.
However, if we assume
that its scale is of the order of 1 TeV and the superpartner masses
are around that scale,
then non-supersymmetric thresholds are not very important for
dimensionless couplings
(which include gauge and Yukawa couplings). Thus, it makes sense to
compute
them and compare them with data in the context of unbroken
supersymmetry.

Threshold corrections appear in the relation between the
running gauge coupling $g_i(\mu)$
of the  low-energy effective field theory
and the string coupling $\gs$ which,
assuming the decoupling of massive modes,
must have the following form:
\be
{16\, \pi^2\over g_{i}^2(\mu)} = k_{i}{16\, \pi^2\over \gs^2} +
 b_{i}\log {M_{s}^2\over \mu^2} +\Delta_{i}\, ,
\label{49}
\ee
where $b_i$ are the usual effective field theory
beta-function coefficients of the
group factor $G_i$, and $k_i$ is the level of the
associated
affine Lie algebra.
The thresholds $\Delta_i$ are due to the
infinite tower of string modes and can be calculated at the level of
string theory. On the other hand, string unification relates the
fundamental string scale
$ M_s\equiv\frac{1}{\sqrt{\alpha'}} $ to the Planck scale
$M_P=\frac{1}{\sqrt{32\, \pi G_N}}$
and to the
string
coupling constant $\gs$ which is associated with the
expectation value
of the dilaton field. At the tree level this relation reads
\begin{equation}
M_s = \gs\,  M_P\label{msmp}\ .
\label{three}
\end{equation}
Given the fact that low-energy data, assuming the minimal
supersymmetric  standard model as
the underlying low-energy field   theory, indicate
gauge  unification at a scale $M_X\sim 2\times 10^{16}$ GeV
\cite{AET}
which is two orders of magnitude less than the Planck scale,
threshold
corrections play a crucial role in string unification.
Their effect has been extensively studied in the literature
\cite{nilles,KPR,DF,ILR} except for the
moduli-dependent universal terms $Y(T,U)$, which
appear in the generic decomposition
\be
\Delta_{i}=\hat \Delta_{i}-k_{i}\,Y \, ,
 \label{52}
\ee
and which have received little attention because they
can be formally reabsorbed into a redefinition of ${\gs}$.
However, such a redefinition alters eq. (\ref{three}) in a
moduli-dependent way and consequently the relation between
string unification scale and Planck mass gets modified.
Following this observation, universal terms were
evaluated explicitly in the context of the symmetric
$Z_2\times Z_2$ orbifold model \cite{pr},
and their effect on on the unification scale of gauge couplings was
shown to consist of  a decrease of the order of $5$ to $10\%$ with
respect
to the case where these corrections are not taken into account.
One of the purposes of the present article is to
extend these results to more general situations. This requires the
computation of one-loop gauge couplings and in particular of
universal threshold corrections in more general string solutions. It
is remarkable that such a computation is possible and exhibits
interesting universality properties as it will appear in the sequel.

There are several procedures for computing the one-loop corrections
to
dimensionless couplings in string theory. The most powerful and
unambiguous one
was described in \cite{kkb,kkc}. It amounts to turning on
gravitational background fields that provide the ground state in
question with a mass gap $\Delta m^2$,
and further background fields (magnetic fields, curvature and
auxiliary $F$ fields) in order to
perform a background-field calculation of the relevant one-loop
corrections.

The above procedure involves the following steps:
\nl
(\romannumeral1)
We first regulate the infra-red by introducing a mass gap in the
relevant ground state. This is done by replacing the flat
four-dimensional conformal field theory with
the wormhole one, $\RR_Q \times SO(3)_{k\over 2}$ \cite{cal, worm}.
The mass gap is given by $\Delta m^2={M_{s}^2 \over 2 (k+2)}$, where
$k$ is a (dimensionless)
non-negative even integer.
\nl
(\romannumeral2)
We then turn on appropriate background fields, which are exact
solutions of the string equations of motion. Such backgrounds include
curvature, magnetic fields and auxiliary $F$ fields\footnote{These
are relevant for the study of the K\"ahler potential renormalization.
For more details see \cite{pet}.}.
\nl
(\romannumeral3)
{We calculate the one-loop vacuum amplitude as a function of these
background fields.
\nl
(\romannumeral4)
We identify these background fields as solutions of the tree-level
effective action. By substituting them into the one-loop effective
action and comparing with the string calculation of the free energy,
we can extract the renormalization constants at one loop.

In the following we will apply the aforementioned techniques to the
calculation of string loop corrections for gauge
couplings. We will derive the full one-loop gauge coupling
for heterotic ground states with at least $N=1$
supersymmetry. We
will in particular obtain an explicit formula for the universal
part of the threshold corrections, and we will
apply this formula
to
show how,
for the whole class  of $N=2$ ground states
that come from two-torus compactification of six-dimensional $N=1$
theories, these thresholds are equal and fully determined as a
consequence of an anomaly-cancellation constraint in six dimensions.
We will also analyse their asymptotic behaviour and singularity
structure. The universal term in particular turns out to be
singularity-free and continuous inside the moduli space, whereas some
second
derivatives with respect to the moduli, such as
$\partial_{T}\, \partial_{\overline{T}}\, Y$, \dots,
are
logarithmically divergent along
enhanced-symmetry planes.
We will then turn to $N=1$ orbifold constructions in four dimensions
and use the above results in order to evaluate exactly the threshold
corrections originated from the $N=2$ sectors.
Although we will not have much to
say concerning a general analytic
formula for the ($N=1$)-sector contributions,
we will present results for two cases, namely the
$Z_3$
and $Z_4$ orbifolds. Here an interesting observation is the breakdown
of
the usual ansatz:
$\Delta_{i}^{N=1}$
cannot in general be decomposed as
$b_i^{N=1}\, \Delta^{N=1}_{\vphantom i}-
k_{i}^{\vphantom N}\,Y^{N=1}_{\vphantom i}$.
Using the above results, we will analyse the effect of the various
thresholds
on the string unification scale, and show that they actually
reduce that scale. Finally, we will clarify the
appearance of the Green--Schwarz term, which is another universal
contribution present in $N=1$ theories. As we will see, this one-loop
correction plays no role for the issue of unification, in contrast
with the universal term $Y$ appearing in the decomposition
(\ref{52}).

We would like to stress here that the study of threshold
corrections is important not only for phenomenological purposes. It
will eventually be useful in the context of string-string dualities
where one expects a deeper understanding of non-perturbative
phenomena \cite{dua}.

\vskip 0.3cm
\setcounter{section}{2}
\setcounter{equation}{0}
\section*{\normalsize{\bf 2. One-loop gauge couplings in heterotic
ground states}}

\noindent
As mentioned previously, an infra-red regulated version of a
heterotic ground state is provided by substituting
four-dimensional flat space with a suitably
chosen conformal field
theory, namely a $(1,0)$ supersymmetric version of
the $\RR_{Q}\times SO(3)_{k\over 2}$ $\s$-model\footnote{The
group $SO(3)$ is required instead of $SU(2)$ for
spin-statistics consistency.} \cite{kkb,kkc}.
This substitution preserves gauge symmetries,
supersymmetry and modular invariance, and introduces curvature
as well as a linear dilaton in the time direction
\be
\Phi={t M_s\over \sqrt{k+2}}\, ,
\ee
necessary for making the total central charge equal to that of flat
space. This amounts to a universal
mass gap for all string excitations (bosonic and fermionic) that
can be read off (in the Euclidean) from the left worldsheet
Hamiltonian
\be
L_{0}=-{1\over 2}+{1\over 4(k+2)}+{p_t^2\over 2}+{j(j+1)\over k+2}+
\cdots\, :
\ee
$\Delta m^2={\mu^2\over 2}$ with $\mu= {M_{s}\over
\sqrt{k+2}}$.

In this geometry, vertices for space-time fields such as
$F_{\mu\nu}^\alpha$
are truly marginal world-sheet operators and therefore deformations
induced
by the associated background fields are exactly calculable. This
allows
in particular for the computation of various one-loop correlators
by inserting the corresponding vertex operators.
For magnetic fields we use
\be
V^{\rm magn}_i\propto \left(J^3 + i : \psi^1 \psi^2 :\right)
\overline{J}_i^{\vphantom m}\, .
\ee
This turns on a magnetic field in the third space direction;
$\overline{J}_i$ is a right-moving affine current in the Cartan of
the $i$th gauge group simple factor
(picking out a single Cartan direction will be enough for our
purposes), and $J^3$ belongs to the $SO(3)_{k\over 2}$ affine Lie
algebra.
There is also a gravitational perturbation generated by
\be
V^{\rm grav}\propto \left(J^3+i:\psi^1\psi^2:\right)\bJ^3\,
{}.
\ee
The currents $J^3$, $\overline{J}^3$ and $\overline{J}_i$ are
normalized so that\footnote{Our normalization is the one widely used
in
the literature; it corresponds to the situation where the highest
root of the algebra has length
squared $\psi ^2=2$.}
\be
J^3(z)\, J^3(0)= {k\over 2z^2}+\cdots\, ,\ \ \bJ^3(\bar z)\,
\bJ^3(0)=
{k\over 2\bar z^2}+\cdots\, ,\ \ \bJ_i(\bar z)\, \bJ_i(0)= {k_i\over
\bar z^2}
+\cdots \, .
\ee
All the above perturbations are products of left times right Abelian
currents
and thus preserve conformal invariance. This implies that the new
backgrounds
satisfy the string equations of motion at tree level to all orders in
the $\ap$ expansion.

Let $\f$ and $\rr$ be constant magnetic and gravitational fields.
The vacuum amplitude at one loop, i.e. the free energy, in the
presence of these backgrounds can be readily calculated by performing
the following Lorentz boost:
\begin{equation}
{\pmatrix{{Q + I^3 \over \sqrt{k/2 +1}} \cr
{\bI^3\over \sqrt{k/2}} \cr
{\bP_{i}\over \sqrt{k_i}}
\cr }}^{\prime}=
\pmatrix{\cosh \phi &\sinh \phi &0 \cr
\sinh \phi &\cosh \phi &0 \cr 0&0&1 \cr }
\pmatrix{1 &0&0 \cr 0 &{\hphantom{-}}\cos \theta&\sin\theta\cr 0
&-\sin\theta&\cos \theta \cr}
\pmatrix{{Q + I^3 \over \sqrt{k/2 +1}} \cr
{\bI^3\over \sqrt{k/2}} \cr
{\bP_{i}\over \sqrt{k_i}}
\cr } \, ,
\label{ecinq}
\end{equation}
where $\f$ and $\rr$ have to be identified with
$\sinh 2\phi \sin \theta$ and
$\sinh 2\phi \cos \theta$, respectively.
Here $I^3,\bI^3$ stand for the zero modes of the respective
$SO(3)_{k\over 2}$ currents,
$Q$ is the zero mode of the $i:\psi^1\psi^2:$ current and $\bP_i$ is
the zero mode of the $\bJ_i$ current.
We assume that the gauge background does not correspond to an
anomalous $U(1)$. This case can also be treated, but is more
complicated\footnote{Since
anomalous $U(1)$'s are broken at scales comparable with the
string scale, their running is irrelevant for low-energy physics.}.
The free energy then reads:
\be
\alpha'^2 F^{\rm \ string}_{\rm one \ loop}=
{1\over 2(2\pi)^4}\int_{\f}{d^2\tau\over (\im)^2} \,
D^{\rm \ string}_{\rm one \ loop}=
{1\over 2(2\pi)^4}\int_{\f}{d^2\tau\over (\im)^2} \,
\left\langle e^{-2\pi\im\, \delta
\left(L_{0}+ \overline{L}_{0}\right)}\right\rangle\, ,
\label{32}
\ee
with
\ba
\delta L_{0}=
\delta \overline{L}_{0}&=&{\sqrt{1+\f^2+\rr^2}-1\over 2}
\left(
{\left(Q+I^3\right)^2\over k+2}
+{1\over \rr^2+\f^2}
\left(\rr{\bI^3\over \sqrt{k}}+\f{\bP_i\over \sqrt{2k_{i}}}\right)^2
\right)\cr
&& +
{Q+I^3\over \sqrt{k+2}}
\left(\rr{\bI^3\over \sqrt{k}}+\f{\bP_i\over \sqrt{2k_{i}}}\right)
\, .
\label{33}
\ea
Expanding to second order in the background fields, we find:
\ba
D^{\rm \ string}_{\rm one \ loop}&=&\langle 1\rangle \cr
&&+
{8\pi^2(\im)^2\rr^2\over k(k+2)}
\left\langle
\left(Q+I^3\right)^2\left(\bI^3\right)^2-
{k\over 8\pi\im}
\left(\left(Q+I^3\right)^2+{k+2\over k}\left(\bI^3\right)^2\right)
\right\rangle \cr
&&+
{4\pi^2(\im)^2\f^2\over k_{i}(k+2)}
\left\langle
\left(Q+I^3\right)^2\bP_i^2-
{k_{i}\over 4\pi\im}
\left(\left(Q+I^3\right)^2+{k+2\over 2 k_i}\bP_i^2\right)
\right\rangle \cr
&&+\cdots
\, ,
\label{34}
\ea
where the dots stand for higher orders in $\f$ and $\rr$.

{}From now on we will assume that our ground state has at least $N=1$
supersymmetry\footnote{The general formula in the absence of
supersymmetry can be found in \cite{kkb}.}.
In such ground states, terms in (\ref{34}) that do not contain the
helicity operator $Q$ vanish because of the presence of the fermionic
zero modes, and terms linear in $Q$ vanish due to rotational
invariance,
$\langle I^3\rangle =0$.
Thus for $N=1$ ground states, (\ref{34}) becomes
\be
D^{\rm \ string}_{\rm one \ loop}=
{8\pi^2(\im)^2\over k+2}\left\langle
Q^2\Bigg( {\rr^2\over k}
\left( \left(\bI^3\right)^2-{k\over 8\pi\im}\right)
+
{\f^2\over 2k_{i}}
\left(\bP_i^2-{k_i\over 4\pi\im}\right)
\Bigg)\right\rangle
+\cdots\, .
\label{35}
\ee
The generic $N=1$ four-dimensional vacuum amplitude has the form
\be
\langle 1\rangle={1\over \im\, |\eta|^4}\sum_{a,b=0,1}
{\vartheta{a\atopwithdelims[]b}\over \eta}\,
C{a\atopwithdelims[]b}\, \Gamma\left({\mu \over M_s}\right)=0\, ,
\label{34a}
\ee
where $C{a\atopwithdelims[]b}$ is the contribution of the internal
conformal field theory, and
\be
\Gamma(x)=-2 x^2
{\partial\over \partial x}
\Big[\sigma(x)-\sigma(2x)\Big]
\ {\rm with}\ \
\sigma(x)={1\over x}\sum_{m,n\in Z}\exp \left(-{\pi  \over \im \,
x^2}|m+n\t|^2\right)
\label{reg}
\ee
at $x=(k+2)^{-{1\over 2}}={\mu / M_s}$
stands for the $SO(3)_{k\over 2}$ partition function
normalized so that
$\lim_{x\to 0}\Gamma(x) = 1$. This
extra factor, which can be
consistently removed,
ensures the convergence of integrals such as those appearing in
(\ref{32}),
at large values of $\im$. Expression (\ref{34a}) allows us to recast
(\ref{35}) as
follows:
\ba
D^{\rm \ string}_{\rm one \ loop}=
-{4\pi i\over k+2}&&\!\!\!\!\!\!\!\!\!{\im\over |\eta|^4}
\sum_{a,b=0,1}
\left\{
{\f^2\over  k_{i}}\,
{\partial_{\t}
\vartheta{a\atopwithdelims[]b}\over\eta}
\left(
\bP_i^2 - {k_{i}\over 4\pi{\im}}\right) C{a\atopwithdelims[]b}\,
\Gamma\left({1\over\sqrt{k+2}}\right)\right.
\cr
&&\!\!\!\!\!\!\!\!\!\!\!-
\left.{\rr^2\over 6 k}\,
{\partial_{\t}
\vartheta{a\atopwithdelims[]b}\over\eta}
C{a\atopwithdelims[]b}\,
\left(\widehat{E}_{2}+{2(k+2)\over i
\pi}\partial_{\tb}\right)\Gamma\left({1\over\sqrt{k+2}}\right)
\right\} +\cdots\, ,
\label{36}
\ea
where $\bP_i^2$ acts as ${i \over \pi}{\partial \over \partial \bar
\tau}$ on the
appropriate subfactor of the 32 right-moving-fermion
contribution, and
\be
\widehat{E}_{2}\equiv {6i\over \pi}\partial_{\tb}\log\left(\im \,
\bar\eta^2\right)
=\bE_2 - {3\over \pi \im} \, ;
\label{377}
\ee
$E_2$ is an Eisenstein holomorphic function
(see (\ref{61b}))
and $\widehat{E}_2$ is modular-covariant of degree
2. Since we are interested in the large-$k$ limit, we can expand
(\ref{36}) in powers of $1/k$. In the next-to-leading order, the
above expression reads:
\ba
D^{\rm \ string}_{\rm one \ loop}=
-{4\pi i\over k}{\im \over |\eta|^4}\,
\Gamma\left({1\over\sqrt{k+2}}\right)
\sum_{a,b=0,1}&&\!\!\!\!\!\!\!\!\!\!\!
\left\{
{\f^2\over  k_{i}}\,
{\partial_{\t}
\vartheta{a\atopwithdelims[]b}\over\eta}
\left(
\bP_i^2 - {k_{i}\over 4\pi{\im}}\right) C{a\atopwithdelims[]b}\,
\left(1-{2\over k}\right)\right.
\cr
&&\!\!\!\!\!\!\!\!\!\!\!-
\left.{\rr^2\over 6 k}\,
{\partial_{\t}
\vartheta{a\atopwithdelims[]b}\over\eta}
\widehat{E}_{2}
\,
C{a\atopwithdelims[]b}+{\cal O}\left({1\over k^2}\right)\right\}
+\cdots\, .
\label{366}
\ea
It deserves stressing here that the
radiative corrections (\ref{366}) include exactly the
back-reaction of the gravitationally coupled fields;
this accounts for the term
$\propto \frac{1}{4 \pi \Im\tau}$, which
is universal and guarantees modular invariance.

In order to determine the string-induced one-loop renormalization
for the gauge couplings, we have to compare the above free energy
with the one that would have been computed in the low-energy
field theory, in the presence of the same backgrounds.
If one normalizes the effective theory generators
such that the highest roots of the group algebra have length
squared equal to $1\, $\footnote{These are the usual normalizations
that
lead in particular to the tree-level relation $M_s={\gs \over
\sqrt{32\, \pi G_N}}$.}, one obtains (more details can be found
in \cite{kkpr}):
\be
\alpha'^2 F^{\rm \, effective}_{\rm \, one \ loop}={1\over k}\left(
-\za\, \f^2
+{\rm higher\ orders\ in\ \f\ and \ \rr}\right)+
{\cal O}\left({1\over k^2}\right)\, ,
\label{46}
\ee
where $\za$ stands for the vector multiplet wave-function
renormalization, as appears in the low-energy effective action
(see (\ref{A1}) and (\ref{A2}))
\be
S_{\, {\rm tree \ \& \ one \ loop}}^{\ \ {\rm gauge \ sector}}
=-\int d^{4}x\,
\sqrt{G} \left(e^{-2\Phi}+\za \right)
\sum_{i,a}{1\over 4 g_{i}^2} \,
F^a_{i\, \mu\nu}\,
F_i^{a\, \mu\nu} \,
{}.
\label{388}
\ee
Comparing eq. (\ref{46})
with eqs. (\ref{32}) and (\ref{366}), and taking into account the
difference between the corresponding normalizations, the
net result for $\za$ reads:
\be
\za \left(\frac{\mu}{M_s}\right)=
{i\over 16\pi^3 k_{i}}
\int_{\cal F}
{d^2\t\over \im}\, {\Gamma(\mu /M_s) \over|\eta|^4}
\sum_{a,b=0,1}
{\partial_{\t}
\vartheta{a\atopwithdelims[]b}\over\eta}
\left(
\bP_i^2 - {k_{i}\over 4\pi{\im}}\right) C{a\atopwithdelims[]b}\, .
\label{47}
\ee
Introducing as usual $\gs = \exp \langle \Phi \rangle$,
we derive from eqs. (\ref{388}) and (\ref{47}) the effective one-loop
string-corrected coupling $g_{i,\,{\rm eff}}$:
\ba
{16\, \pi^2\over g_{i,\,{\rm eff} }^2}&=&k_{i}{16\, \pi^2\over
\gs^2}+
16\, \pi^2 k_{i}\,\za \left(\frac{\mu}{M_s}\right) \cr
&=&k_{i}{16\, \pi^2\over \gs^2}+{i\over \pi}
\int_{\cal F}
{d^2\t\over \im}\, {\Gamma(\mu /M_s) \over|\eta|^4}
\sum_{a,b=0,1}
{\partial_{\t}
\vartheta{a\atopwithdelims[]b}\over\eta}
\left(
\bP_i^2 - {k_{i}\over 4\pi{\im}}\right) C{a\atopwithdelims[]b}\,
.\label{48}
\ea
Equation (\ref{48}) has been obtained by using an explicitly
infra-red-regulated string loop
amplitude. However, it is important to stress that the final
relation between the running gauge couplings of the low-energy field
theory and the string coupling should not depend on how the
infra-red has been regulated;
put differently, this relation should not depend on the function
$\Gamma(\frac{\mu}{M_s})$. In order to show this property,
and eventually establish the expression for the running low-energy
gauge couplings,
we first isolate the
contribution of the massless
states responsible for the non-trivial infra-red behaviour of the
integral in
(\ref{47}): we rewrite (\ref{48}) in a form
where we subtract and add back a $b_i\int_{\cal
F}\frac{d^2\tau}
{\Im \tau}\Gamma(\frac{\mu}{M_s})$ term, a manipulation perfectly
well
defined
thanks to
the presence of the regulator $\Gamma(\frac{\mu}{M_s})$.
Here $b_i$ are the full beta-function coefficients for the group
factor
$G_i$:
\be
b_{i}=
\lim_{\im\to\infty}{i\over \pi}\, {1 \over|\eta|^4}
\sum_{a,b=0,1}
{\partial_{\t}
\vartheta{a\atopwithdelims[]b}\over\eta}
\left(
\bP_i^2 - {k_{i}\over 4\pi{\im}}\right) C{a\atopwithdelims[]b}\, .
\label{50}
\ee
Using the result
\begin{equation}
\int_{\cal F}\frac{d^2\tau}
{\Im \tau}\, \Gamma\left(\frac{\mu}{M_s}\right) =
\log\frac{M_s^2}{\mu^2}
+ \log\frac{2e^{\gamma+3}}{\pi\sqrt{27}}+{\cal O}
\left(\frac{\mu}{M_s}\right)\, ,
                         \label{thirteen}
\end{equation}
and taking the limit $\mu\to 0$ in the remaining
integral
of (\ref{48})
since it does not suffer any longer
from divergences at $\Im\tau\to\infty\,$ leads to:
\ba
{16\, \pi^2\over g_{i,\,{\rm eff} }^2}&=&k_{i}{16\, \pi^2\over
\gs^2}+
b_{i}\log {M_{s}^2\over \mu^2} +
b_{i}\log {2\,e^{\gamma +3}\over \pi\sqrt{27}}
\cr
&&+\int_{\cal F}{d^2\t\over \im}
\left(
{i\over \pi}
\, {1 \over|\eta|^4}
\sum_{a,b=0,1}
{\partial_{\t}
\vartheta{a\atopwithdelims[]b}\over\eta}
\left(
\bP_i^2 - {k_{i}\over 4\pi{\im}}\right) C{a\atopwithdelims[]b}-
b_{i}
\right)
\, .\label{KKB}
\ea

We can determine the running gauge couplings of the
low-energy effective field theory by identifying the above
string theory one-loop corrected coupling
$g_{i,\,{\rm eff}}$ with the corresponding field theory
one-loop gauge coupling,  regulated
in the infra-red in a similar fashion as the string theory.
The effective field theory has also to be supplied
with an ultraviolet cut-off. If we use dimensional regularization,
we obtain the following field theory one-loop corrected coupling:
\begin{equation}
\left.{16\, \pi^2\over g_{i,\,{\rm eff} }^2}
\right\vert_{\rm \ field \ theory}=
\frac{16\, \pi^2}{g_{i,\, {\rm bare}}^2} +
b_i\, (4\pi)^\epsilon\int_0^\infty\frac{dt}{t^{1-\epsilon}}\,
\Gamma_{FT}\left(\frac{\mu}{\sqrt{\pi} M}\right)\, ,
                     \label{six}
\end{equation}
where
$M$ is an arbitrary mass
scale,
and $\Gamma_{FT}$ is the field theory counterpart of the string
infra-red regulator, obtained by dropping all winding modes
\footnote{The extra $\sqrt{\pi}$ in the
argument of $\Gamma_{FT}$
accounts for the identification of the (dimensionless in the above
convention)
Schwinger proper-time parameter $t$ with $\pi\Im\tau\,$.}
in (\ref{reg}). On the
other hand, one knows
that in the $\overline{DR}$ scheme the relation between the field
theory bare
and running coupling is
\begin{equation}
\frac{16\, \pi^2}{g_{i,\, {\rm bare}}^2} =
\frac{16\, \pi^2}{g_{i}^2(\mu)} -
b_i \, (4\pi)^\epsilon\int_0^\infty\frac{dt}
{t^{1-\epsilon}}\, e^{-t\frac{\mu^2}{M^2}}\, .
                      \label{seven}
\end{equation}
Plugging (\ref{seven}) into (\ref{six}) and performing the resulting
integral in the limit $\mu,\,  \epsilon \to 0$, leads to the
following expression for the field theory one-loop corrected
coupling:
\begin{equation}
\left.{16\, \pi^2\over g_{i,\,{\rm eff} }^2}
\right\vert_{\rm \ field \ theory}^{\ \overline{DR}}=
\frac{16\, \pi^2}{g_{i}^2(\mu)} +  b_i\, (2\gamma+2)\, ;
   \label{KK}
\end{equation}
identifying the latter
with
(\ref{KKB}), we finally obtain \cite{pr}
the already anticipated eq. (\ref{49})
with
\be
\Delta_{i}=\int_{\cal F}{d^2\t\over \im}
\left(
{i\over \pi}
\, {1 \over|\eta|^4}
\sum_{a,b=0,1}
{\partial_{\t}
\vartheta{a\atopwithdelims[]b}\over\eta}
\left(
\bP_i^2 - {k_{i}\over 4\pi{\im}}\right) C{a\atopwithdelims[]b}-
b_{i}
\right)
+b_{i}\log {2\,e^{1-\gamma}\over \pi\sqrt{27}}\, .
\label{51}
\ee
These are the full threshold corrections in the $\overline{DR}$
scheme.
As advertised previously, this expression no longer depends
on the infra-red regularization prescription.
This result could have been anticipated as a consequence of the
cancellation of the infra-red divergences between the fundamental and
the effective theory since they have the same massless spectrum.
However it could only be proved
\cite{pr} in the presence of a consistent infra-red regulator,
similar in both theories. Moreover, it is important to emphasize that
(\ref{51}) contains rigorously all universal terms that were missing
in
previous approaches
\cite{kaplu, dkl}
and that we will be analysing in the sequel.

We would like to stress that the computation we presented here was
performed in the {\it dilaton frame} for {\it both} string theory and
the
effective field theory in which we have used a moduli-dependent
convention for
the masses. To put it differently, the kinetic terms of
the various fields are normalized to one, and this is a physical
basis.
In this frame, the string perturbative expansion appears as a power
series
with respect to the coupling
$\gs = \exp \langle \Phi \rangle$,
which is a well defined parameter that remains unaltered
at any order of the expansion. Therefore, it provides an
unambiguous control of the latter. Moreover, as long as the
string ground state possesses at least $N=1$ supersymmetry, the
Planck scale $M_P$ does not receive any correction in
perturbation theory \cite{kkpr},
which means that the tree-level relation (\ref{three}) holds to all
orders.
This property plays an important role in the low-energy unification
of the effective couplings \cite{pr}.

Instead of the dilaton frame,
one could use the $S$-frame
in the effective supergravity.
In that case, the expansion parameter
is ${1\over \Im S}$, and this turns out to be convenient for
analysing the holomorphicity and duality properties which are
somehow
obscured in the dilaton frame, and consequently the issue
of supersymmetry.
On the other hand, this parameter must be redefined at each order of
the perturbative expansion, as a consequence of the
antisymmetric-moduli
mixing due to the Green--Schwarz term \cite{dfkz}.
This term changes the definition of the axion and, by supersymmetry,
that
of the dilaton.
Thus, terms of order $n$ in the $S$-frame get contributions from all
loop orders up to the $n$th.
Furthermore, a new universal-threshold-like correction appears now
along with the thresholds (\ref{51}), which is again a remnant of the
ten-dimensional Green--Schwarz term \cite{dfkz, kl}.
This term is only present in $N=1$ models and is moduli-dependent
in contrast (see section 4) with the ($N=1$)-sector contribution of
the corrections
(\ref{51}). This extra moduli dependence, which enters from the
effective field theory matching in the $S$-frame, is responsible for
the modification of the analytic
properties in this frame.
The appearance of the Green--Schwarz
term
in the $S$-frame is summarized in appendix A.

As a final remark, we would like to comment on the structure of the
threshold corrections as they appear in the dilaton frame, which is
the frame that we will be using in our subsequent computations.
Part of the thresholds (\ref{51}) is universal and this enables us
to split (\ref{51})
according to eq. (\ref{52}).
As we have already mentioned,
the universal piece $Y$ in (\ref{52}) contains, among other things,
contributions
from
the universal sector (gravity in particular).
Such contributions are not taken into account in grand unified
theories
while in string theory they are well defined and calculable
quantities.
Moreover, $Y$ is infra-red-finite, which in particular means that it
is continuous and remains
finite when extra states become massless at some special values of
the moduli. Thus $Y$ is a finite correction to the ``bare" string
coupling $\gs$, and we can write (\ref{49}) as
\be
{16\, \pi^2\over g_{i}^2(\mu)}=k_{i}{16\, \pi^2\over g_{\rm
renorm}^2}+
b_{i}\log {M_{s}^2\over \mu^2} +\hat \Delta_{i}\, ,
\label{53}
\ee
where we have defined a ``renormalized" string coupling by \cite{pr}
\be
g_{\rm renorm}^2={\gs^2\over 1-{Y\over 16\, \pi^2}\,\gs^2 }\, .
\label{55}
\ee
Of course, such a coupling is meaningful provided it appears as the
natural
expansion parameter in several amplitudes that are relevant for the
low-energy string physics. In general, this might not be the case as
a consequence
of some arbitrariness in the decomposition (\ref{52}). Examples of
this
kind arise in $N=1 $ models (see the $Z_4$ orbifold in section 4) as
well as in certain $N=2$ constructions \cite{deco}.
It is important to keep in mind that this ``renormalized"
string coupling is defined here in a {\it moduli-dependent} way.
This moduli dependence affects the string unification
\cite{pr}. Indeed, as we will see in the sequel, when proper
unification of the couplings appears, namely when $\hat\Delta_i$ can
be written
as $b_i\Delta$, their common value at the
unification scale is $g_{\rm renorm}$, which plays therefore the role
of a phenomenological parameter.
Moreover,
the unification scale
turns out to be proportional to $M_s$. The latter can
be
expressed in terms of
the ``low-energy" parameters
$g_{\rm renorm}$
and $M_P$, by using eq. (\ref{three})
and its non-renormalization property \cite{kkpr}, as well as
(\ref{55}):
\be
M_s={M_{P}
\, g_{\rm renorm} \over \sqrt{1 + \frac{Y}{16\, \pi^2}\, g_{\rm
renorm}^2}}
\, ;
\label{A16}
\ee
this involves the moduli-dependent function $Y$.
As is shown in appendix A, relation (\ref{A16}) holds also in
the $S$-frame where, at one loop,
$ g_{\rm renorm}^{\, -2}=\Im S + {\Delta^{GS} \over 16\, \pi^2}$
with $\Delta^{GS}$ the Green--Schwarz term
(see eq. (\ref{grenS})). Despite the presence of the moduli-dependent
universal function $\Delta^{GS}$, the string scale $M_s$, and
consequently the unification scale, are only affected by the
universal threshold $Y$.
\vskip 0.3cm
\setcounter{section}{3}
\setcounter{equation}{0}
\section*{\normalsize{\bf 3. Universal thresholds for a class of
$N=2$ theories}}

\noindent
Let us now concentrate on  $N=2$ ground states.
We will focus on models
that come from toroidal compactification of generic
six-dimensional
$N=1$ string  theories. There are of course more general $N=2$ models
in four dimensions that cannot
be viewed as toroidal compactifications of a six-dimensional theory
\cite{deco}. These will be dealt with in detail in a separate
publication. In the cases at hand, however, there is a universal
two-torus, which provides the (perturbative) central charges of the
$N=2$ algebra. Therefore, (\ref{51})  becomes
\be
\Delta_{i}=\int_{\cal F}{d^2\t\over \im}
\Bigg({\Gamma_{2,2}\left(T,U,\bT,\bU\right)\over \bar \eta^{24}}
\left(\bP_{i}^2-{k_{i}\over 4\pi\im}
\right)\overline{\Omega}-b_{i}
\Bigg)+b_{i}\log {2\, e^{1-\gamma}\over \pi\sqrt{27}}\, ,
\label{56}
\ee
where $T$ and $U$ are the complex moduli of the two-torus,
$\overline{\Omega}$ is an antiholomorphic function
and
\ba
\Gamma_{2,2}\left(T,U,\bT,\bU\right)=
\sum_{{\mbox{\footnotesize\bf m}},{\mbox{\footnotesize\bf n}}\in Z}
\exp
\bigg(\!\!\!\!\!\!\!\!\!\!&&
2\pi i\t\left(
m_{1}\, n^{1}+m_{2}\, n^{2}
\right)\cr &&-
{\pi\im\over \Im T \Im U}
\left|T n^{1} + TUn^{2}+Um_1-m_2\right|^2
\bigg)\, .
\label{60}
\ea
{}From (\ref{56}), we observe that the function
\be
\bF_{i}={1\over \bar \eta^{24}}
\left(\bP_{i}^2-{k_{i}\over 4\pi\im}
\right)
\overline{\Omega}
\label{57}
\ee
is modular invariant.
Consider the associated function that appears in the
$R^2$-term renormalization (see eq. (\ref{366}) or ref. \cite{antg}
for more details),
\be
\bF_{\rm grav}=
{\widehat{E}_2\over 12}\, {\bOmega \over \bar \eta^{24}}
={1\over \bar \eta^{24}}\left({i\over
\pi}\partial_{\tb}\log\bar\eta-{1\over 4\pi\im}\right)\bOmega
\, ,
\label{58}
\ee
which is also modular invariant, and eventually leads to the
gravitational anomaly.
The difference $\bF_{i}-k_{i}\, \bF_{\rm grav}$ is an antiholomorphic
function, which is modular invariant. It has at most a simple pole at
$\t\to i\infty$ (associated with the heterotic unphysical tachyon)
and is finite inside the moduli space of the torus. This implies that
\be
\bF_{i}=k_{i}\, \bF_{\rm grav}+A_i\, \bj(\tb)+B_i\, ,
\label{59}
\ee
where $A_i$ and $B_i$ are constants to be determined, and
$j(\t)={1\over q}+744+{\cal O}(q)$,
$q=\exp(2\pi i\t)$ is the standard $j$-function.
The modular invariance of $\bF_{\rm grav}$ implies that $\bOmega$ is
a modular form of weight 10, which is finite inside the moduli
space.
This property fixes
\be
\bOmega =\xi \,  \bE_{4}\, \bE_{6}\, ,
\label{61}
\ee
where $E_{2n}$ is the $n$th Eisenstein series:
\be
E_{2}=
{12\over i \pi}\partial_{\t}\log \eta
=1-24\sum_{n=1}^{\infty}{n\, q^n\over 1-q^n}
\, ,
\label{61b}
\ee
\be
E_{4}=
{1 \over 2}\left(
{\vartheta}_2^8+
{\vartheta}_3^8+
{\vartheta}_4^8
\right)
=1+240\sum_{n=1}^{\infty}{n^3q^n\over 1-q^n}
\, ,
\ee
\be
E_{6}=
\frac{1}{2}
\left({\vartheta}_2^4 + {\vartheta}_3^4\right)
\left({\vartheta}_3^4 + {\vartheta}_4^4\right)
\left({\vartheta}_4^4 - {\vartheta}_2^4\right)
=1-504\sum_{n=1}^{\infty}{n^5q^n\over 1-q^n}
\, .
\ee
Putting everything together in (\ref{56}) we obtain:
\be
\Delta_{i}=\int_{\cal F}{d^2\t\over \im}\,
\Bigg(\Gamma_{2,2}\left(T,U,\bT,\bU\right)
\left({\xi k_{i}\over 12}{\widehat{E}_{2}\,
\bE_{4}\, \bE_{6}\over \bar \eta^{24}}+A_i \, \bj+B_i\right)-b_{i}
\Bigg)+b_{i}\log {2\, e^{1-\gamma}\over \pi\sqrt{27}}
\, .
\label{62}
\ee

There are two constraints that allow us to fix the constants $A_i,
B_i$.
The  first is that the $1/\bar q$ pole is absent from the group
trace,
which gives
\be
A_i=-{\xi k_{i}\over 12}\, .
\label{64}\ee
The second is (\ref{50}), which implies
\be
744\, A_i+B_i-b_{i}+ k_i\,  b_{\rm grav} = 0\, ,
\label{63}\ee
where the constant term in the large-$\im$
expansion of
$\bF_{\rm grav}$
\be
b_{\rm grav} =  \lim_{\im \to\infty}\left(\bF_{\rm grav}
-{\xi \over 12}{1 \over \bar q}\right)
= -22\,\xi
\label{bla}
\ee
is the gravitational anomaly in units where  a hypermultiplet
contributes
${1\over 12}$ \cite{antg}.
Plugging (\ref{64})--(\ref{bla}) in (\ref{62}), we finally obtain:
\ba
\Delta_{i}&=&
b_{i}\left(
\log {2\, e^{1-\gamma}\over \pi\sqrt{27}}+\int_{\cal F}{d^2\t\over
\im}\, \Big(\Gamma_{2,2}\left(T,U,\bT,\bU\right)-1\Big)
\right)\cr
&&+{\xi k_{i}\over 12}\int_{\cal F}{d^2\t\over \im}\,
\Gamma_{2,2}\left(T,U,\bT,\bU\right)
\left({\widehat{E}_{2}\, \bE_{4}\, \bE_{6}\over \bar
\eta^{24}}-\bj+1008\right)
\, .
\label{65}
\ea
The first integral in (\ref{65}) was computed explicitly in
\cite{dkl} and recently generalized in \cite{hm}:
\be
\int_{\cal F}{d^2\t\over \im}
\Big(\Gamma_{2,2}\left(T,U,\bT,\bU\right)-1\Big)=
-\log\left(\big|\eta(T)\big|^4 \big|\eta(U)\big|^4
\Im T  \Im U\right)-\log{8\pi \, e^{1-\gamma}\over \sqrt{27}}\, .
\label{66}
\ee
Therefore, as advertised above, we can write\footnote{This is not
true
for more general $N=2$ ground states (see \cite{deco}).}
\be
\Delta_{i}=b_{i}^{\vphantom N}\,
\Delta-k_{i}^{\vphantom N} \, Y\, ,
\label{66a}
\ee
with
\be
\Delta=-\log\left(4\pi^2 \big|\eta(T)\big|^4 \big|\eta(U)\big|^4
\Im T  \Im U\right)
\label{67}
\ee
and
\be
Y=-{\xi \over 12}\int_{\cal F}{d^2\t\over \im}\,
\Gamma_{2,2}\left(T,U,\bT,\bU\right)\, \Bigg(
\left(\bE_{2}-{3\over \pi\im}\right){\bE_{4}\, \bE_{6}\over \bar
\eta^{24}}-\bj+1008
\Bigg)
\label{68}
\ee
(we have used eq. (\ref{377})).
This form of the universal term was determined for the case of
$Z_{2}\times Z_{2}$ orbifolds in \cite{pr} and further discussed
in \cite{kkpr}.

The coefficient $\xi$ can be related to the number of massless
vector multiplets $N_{V}$ and hypermultiplets $N_{H}$ via the
gravitational
anomaly ($b_{\rm grav}$), which
can also be computed from the low-energy theory of massless states.
In units where a scalar contributes $1$,
the graviton contributes $212$, the antisymmetric tensor $91$, the
gravitino $-{233\over 4}$, a vector~$-13$ and a Majorana fermion
$7\over 4$; therefore
the $N=2$ supergravity multiplet contributes $212-2{233\over
4}-13={165\over 2}$, the tensor multiplet contributes $-13+2{7\over
4}+1+91={165\over 2}$,
a vector multiplet $-13+2{7\over 4}+2=-{15\over 2}$ and a
hypermultiplet
$2{7\over 4}+4={15\over 2}$.
Thus in the units of (\ref{bla}),
\be
b_{\rm grav} = {22 - N_V +N_H\over 12}\, ,
\ee
and hence
\be
\xi=-{1\over 264}\left(22-N_{V}+N_{H}\right)\, .
\label{70}
\ee
For the models at hand we can go even further and completely
determine
$\xi$. One
can indeed show that $N_{H}-N_{V}$ is a {\it universal constant}
for the whole class of four-dimensional $N=2$ models obtained
by toroidal compactification of {\it any} $N=1$ ground state
in six dimensions. The argument is the following. From the
six-dimensional point of view,
the models at hand must obey an anomaly-cancellation
constraint,  which reads:~$\left. N_{H}-N_{V}\right|_{\rm six \
dim}=244$,
and does not depend on the kind of compactification that has been
performed from ten to six
dimensions\footnote{Actually, this constraint, which ensures that
$\Tr R^4$ vanishes, holds even when there occurs a symmetry
enhancement originated from non-perturbative effects, provided the
number of tensor multiplets remains $N_T = 1$. Note that this
six-dimensional anomaly-cancellation constraint is also used
in \cite{uu}, in relation to four-dimensional quantities.}
\cite{sch}.
After two-torus compactification to four dimensions,
two extra $U(1)$'s appear, leading to the relation
\be
N_{H}-N_{V}=242
\label{anom}
\ee
between the
numbers of vector multiplets and hypermultiplets.
In turn, eq. (\ref{70}) implies that for this class of
ground states $\xi=-1$.
As a consequence,
all $N=2$ models under consideration have equal
universal thresholds, given by
\be
Y\left(T,U,\bT,\bU\right)={1\over 12}\int_{\cal F}{d^2\t\over \im}\,
\Gamma_{2,2}\left(T,U,\bT,\bU\right)\,
\Bigg(\left(\bE_2-{3\over \pi\im}\right){\bE_4\, \bE_6\over \bar
\eta^
{24}}-\bj+1008
\Bigg)\, .
\label{71}
\ee
As an example, consider
the case of the $Z_{2}$ orbifold, where we have a gauge group
$E_{8}\times
E_7\times SU(2)\times U(1)^2$ and thus $N_V=386$.
The number of massless hypermultiplets is $N_H=628$.
Using these numbers in (\ref{70}) we obtain indeed $\xi=-1$. As
expected by supersymmetry, the corresponding universal threshold
(\ref{71}) is
twice as big as a single-plane contribution of the symmetric
$Z_{2}\times Z_{2}$ orbifold analysed in
\cite{pr}\footnote{The universal threshold computed in \cite{pr}
corresponds to the three-plane contribution of the $Z_2 \times Z_2$
model with $T_i=T$ and $U_i = U$ $\forall i$. A factor $2/3$ is
therefore needed to recover (\ref{71}).}.

Expression (\ref{71}) can be further simplified if one uses a
generalization of (\ref{66}), valid for more general
modular-invariant functions, to integrate the last terms:
\ba
Y\left(T,U,\bT,\bU\right)&=&{1\over 12}\int_{\cal F}{d^2\t\over
\im}\,
\Gamma_{2,2}\left(T,U,\bT,\bU\right)\,
\Bigg(\left(\bE_2-{3\over \pi\im}\right){\bE_4\, \bE_6\over \bar
\eta^
{24}}+264
\Bigg)\cr
&&+{1\over 3}
\log\big\vert j(T)-j(U)\big\vert
\, .
\label{logd}
\ea
Using the method of orbits of the modular group, the remaining
integral in (\ref{logd}) can also be reduced to a multiple series
expansion (see \cite{hm}). This is detailed in appendix B, where we
use a product representation of $j(T)-j(U)$ to make more transparent
the cancellation of the logarithmic divergences occurring in both
terms of (\ref{logd}) when $T\to U$ as a consequence of the
appearance of extra massless states. It is actually shown that
$Y$ is {\it finite and continuous} inside the whole moduli space.
We also present
in this appendix
numerical evaluation and plots as well as
various large-moduli behaviours of (\ref{71}); the latter
have some relevance in the context of the decompactification problem
(see \cite{deco}).

As far as the issue of unification is concerned, several observations
are in order.
Although $N=2$ models are not directly relevant for phenomenology, it
is nevertheless interesting to note that universal thresholds always
decrease
the unification scale. Indeed, by using eqs. (\ref{53}) and
(\ref{66a}),
it appears that unification of all couplings takes place (in the
$\overline{DR}$ scheme) at a scale
\begin{equation}
M_U=  M_P\,g_U \, e^{\Delta \over 2} \, \frac{1}
{\sqrt{1 + \frac{Y}{16\, \pi^2}\, g_U^2}}\, ,
\label{dmu}
\end{equation}
where
\be
g_U\equiv g_\alpha(M_U)=
{\gs \over \sqrt{1-\frac{Y}{16\,\pi^2}\, \gs^2}}
\ee
for any group factor
(this is the renormalized coupling introduced in (\ref{55})),
$\Delta$ and $Y$ are given by eqs. (\ref{67}) and (\ref{logd})
respectively, and we
have
used
explicitly (\ref{three}) in order to express the unification scale in
terms of the effective field theory parameters $M_P$ and $g_U$.
The last factor in (\ref{dmu}) is due to the existence
of the universal terms which lead to a shift of the dilaton field in
order to reabsorb the universal contributions into the string
coupling.
It is interesting to observe that\footnote{We
consider here the case $\Re T = \Re U = 0$, and we
parametrize it as usually:
$
\Im T= R_1\, R_2
$
and
$
\Im U= R_2 / R_1
$.}
$Y(R_1,R_2) > 0$ (see appendix B).
Therefore, this extra
factor
always gives  a lower unification scale with respect
to the case where these terms are neglected.
On the other hand the first
factor in (\ref{dmu}) monotonically increases for  radii moving away
from
the self-dual point, while the second one monotonically decreases.
Following \cite{pr} we conclude that the minimum unification scale is
reached at the self-dual point $R_1 = R_2 =  1$ with the value
\begin{equation}
M_U^{\scriptstyle\rm min}
\approx 5.56\times 10^{17}\times g_U\times
\frac{1}
{\sqrt{1 + 0.15 \times g_U^2}}
\, {\rm GeV}\, .
\label{dmumin}
\end{equation}
The last factor in this expansion represents the effect of the
universal thresholds. Note that results (\ref{dmu}) and
(\ref{dmumin})
are valid for the whole class of $N=2$ models that were analysed
here above.

Besides the relevance that the universal contributions
$Y$
might have in the framework of string unification, we should mention
that they are also related to the one-loop correction of the K\"ahler
potential for the moduli fields \cite {dfkz, kl, ant, pet}
(see also appendix A, eqs. (\ref{A7}) and (\ref{A15})), as is
expected from
supersymmetry:
\begin{equation}
K^{(1)}\left(T,U,\overline{T},\overline{
U}\right)=-{1\over 16\, \pi^2}\,
Y\left(T,U,\overline{T},\overline{U}\right) +
\kappa\Big(T,U\Big) +
{\bar\kappa}\left(\overline{T},\overline{U}\right)\, ;
\label{kpot}
\end{equation}
here $\kappa\big(T,U\big)$ is an analytic function which is
irrelevant in the determination of the K\"ahler metric but plays a
role in the duality covariance of $K^{(1)}$. By using the identity
\begin{equation}
{1 \over (\Im T)^2}\,
{\partial ^2\over \partial  \tau \partial \bar\tau}
\left(\Im  \tau \,
\Gamma_{2,2}
\right)
={1\over \Im \tau }\,
{\partial ^2\over \partial T \partial \bT}\,
\Gamma_{2,2}
\end{equation}
in eqs. (\ref{71}) and (\ref{kpot}),
it is easy to show that
\ba
{\partial ^2 K^{(1)}
\over \partial T \partial \bT}&=&
-{1\over 128 \, \pi^2 \left( \Im T\right)^2}
\int_{\cal F}
{d^2\tau \over (\Im \tau)^2}
\,
{i\over  \pi}
\, {\partial \over \partial \bar \tau}
\big(
\Im \tau\,
\Gamma_{2,2}
\big)
\,
{\bE_4 \, \bE_6
\over {\bar \eta}^{24}}\cr
&&-
{1\over 192 \, \pi^2 \left( \Im T\right)^2}
\int_{\cal F}
{d^2\tau }
\,{\partial \over \partial  \tau}\,
G(\tau, \bar \tau)\, ,
\label{dkpot}
\ea
where
\be
G(\tau, \bar \tau)=\left(
{i \over 2}\, \Gamma_{2,2}
+\im  {\partial \over \partial \bar \tau}
\Gamma_{2,2}
\right)\left(
{\bE_2 \,\bE_4 \, \bE_6
\over {\bar \eta}^{24}}
-\bj+1008
-{3\over \pi\im}\, {\bE_4\, \bE_6\over \bar
\eta^{24}}
\right)
\, .
\label{G}
\ee
The first term in (\ref{dkpot}) is indeed the one-loop correction to
the K\"ahler metric
$K^{\, (1)}_{T \bT}$
as it appears in \cite{antgg}\footnote{To make contact with this
reference,
one can use the following identity:
$$
{E_4 \, E_6\over\eta ^{24}}=\big(j(i)-j(\tau)\big)
\left({\partial \log j\over \partial \log q}\right)^{-1}
\, .$$
} or \cite{pet}. The second one is a boundary term that vanishes for
generic values of the moduli. However, when $T \sim U$,
$\Gamma_{2,2}\sim 1 + \bar q + \cdots  $, and
this term might develop finite or even $\delta$-function
contributions.
Actually, $\delta$-functions are originated around $T=U$ as
\ba
\lim_{\im \to \infty}\im \left( {\partial \over \partial \bar \tau}
\Gamma_{2,2}\right) {1\over \bar q}&=&
-4\pi i \lim_{\im \to \infty}\im \,  \exp \left(-\pi\im {\vert T-U
\vert^2\over
2\Im T \Im U}\right)\cr
&=& -8\pi i \, (\Im T)^2 \, \delta^2 (T-U)
\, .
\ea
It is nice to observe that these contributions are avoided for the
same reason that $Y$
remains free of infra-red divergences when the gauge symmetry gets
enlarged: the absence of poles in ${\bE_2 \,\bE_4 \, \bE_6
\over {\bar \eta}^{24}}
-\bj+1008$. However, the term
$-{3\over \pi\im}\, {\bE_4\, \bE_6\over \bar
\eta^{24}}$
in (\ref{G})
will generate finite boundary contributions at $T=U$, which
will be further enhanced when $T=U=i$ and $T=U=\rho=\exp {2\pi i\over
3}$.
More precisely we have
\be
-{1\over 192 \, \pi^2 \left( \Im T\right)^2}
\int_{\cal F}
{d^2\tau }
\,{\partial \over \partial  \tau}\,
G(\tau, \bar \tau)=-{\l \over 64 \, \pi^2 \left( \Im T\right)^2}
\ee
with $\l=6$ for $T=U=\rho$, $\l=4$ for $T=U=i$, $\l=2$ for generic
$T=U$,
and $\l=0$ elsewhere.

Finally, by using expression (\ref{Bexp}) it is easy to show that
$Y$
is continuous at $T=U$ as long as the moduli remain finite. On the
other hand,
$\partial_{T}\, \partial_{\overline{T}}\, Y$ develops logarithmic
singularities:
\be
\partial_{T}\, \partial_{\overline{T}}\, Y\sim
-{1\over (\Im T)^2}\, \log\vert T-U\vert
+{\rm \ regular \ terms,}
\label{dY}
\ee
around $T\sim U$, as is expected from general arguments. Those
singularities are responsible for
non-trivial monodromy properties of the prepotential in these $N=2$
models \cite{antgg, lust, hm}. In connection with these properties, a
comment is in order here: our present treatment implies that the
results
of \cite{hm} for the case of vanishing Wilson lines provide the
prepotential
for all $N=2$ models that are toroidal compactifications of $N=1$
six-dimensional theories. Furthermore,
this holds also for the special classes of Wilson lines dealt
with in \cite{hm}.
\vskip 0.3cm
\setcounter{section}{4}
\setcounter{equation}{0}
\section*{\normalsize{\bf 4. The case of $N=1$ orbifolds}}

\noindent
We come now to the $N=1$ heterotic compactifications on orbifolds
${\bf T}^6 /G$.  We will restrict to the case where $G$ is Abelian.
Although there is not any fundamental obstruction with the
non-Abelian
situation, the computation is expected to be more
complicated.
In a generic $N=1$ orbifold compactification,
both $N=1$ and $N=2$
supersymmetric sectors contribute to the gauge coupling
renormalization.
This allows us to express the thresholds (\ref{51}),
which appear in eq. (\ref{49}), as
\ba
\Delta_{i}^{\vphantom N}&=&
\Delta_{i}^{N=1}+\Delta_{i}^{N=2}
\cr
&=&\sum_{N=1,2}\int_{\cal F}{d^2\t\over \im}
\left(
{i\over \pi}
\, {1 \over|\eta|^4}
\sum_{a,b=0,1}
{\partial_{\t}
\vartheta{a\atopwithdelims[]b}\over\eta}
\left(
\bP_i^2 - {k_{i}\over 4\pi{\im}}\right)
C^N_{\vphantom i}{a\atopwithdelims[]b}-
b_{i}^N
\right)\cr
&&+b_{i}^N\log {2\,e^{1-\gamma}\over \pi\sqrt{27}}\, ,
\label{511}
\ea
where $b_{i}^{N=1,2}$ are the contributions originated from
$N=1,2$ supersymmetric sectors and
$b_{i}^{\vphantom N}=b_{i}^{N=1}+b_{i}^{N=2}$ are the full
beta-function coefficients of the $N=1$ model.

Let us focus for the moment on the ($N=2$)-sector
contributions,
which exhaust all the dependence of the thresholds on the
untwisted moduli of the torus ${\bf T}^6 /G$. Such sectors are
present
when some twists $g \in G$ have unit eigenvalues, leaving therefore
unrotated a complex plane of ${\bf T}^6$. As was described in
\cite{dkl}, these twists form a disjoint union
$\bigcup_{\a} G_{\a} \subset G$: they only share the identity
corresponding to the $N=4$ sector. Each $G_{\a}$ is the little group
of a given complex plane of ${\bf T}^6$, which needs not be the same
for all $G_{\a}$'s. Furthermore, the subset of sectors generated by
twists $g \in G_{\a}$ appears actually as the set of all twisted
sectors
that define an orbifold model on ${\bf T}^6 /G_{\a}$.
Thus we conclude that
\be
\Delta_{i}^{N=2}\left[{\bf T}^6 /G
\right]=\sum_{\a}{\left\vert G_{\alpha}\right\vert \over \vert
G\vert}
\, \Delta_{i}^{\vphantom N}
\left[{\bf T}^6 /G_{\a}
\right]\, .
\label{513}
\ee
On the other hand, the orbifold model on ${\bf T}^6 /G_{\a}$ has
$N=2$ supersymmetry and can be viewed as a two-torus compactification
of a $N=1$ supersymmetric model in six dimensions, the two-torus
being the complex plane of ${\bf T}^6$ left invariant under $G_{\a}$.
Hence it belongs to the class of models that have been studied in
section 3, leading automatically to the following result:
\be
\Delta_{i}^{N=2}=-
\sum_{\a}b_i^{\a}\,  {\left\vert G_{\alpha}\right\vert \over \vert
G\vert} \,
 \log\left(4\pi^2 \big|\eta(T_{\a})\big|^4 \big|\eta(U_{\a})\big|^4
\Im T_{\a}  \Im U_{\a}\right)-k_i^{\vphantom N}\,
Y_{\vphantom i}^{N=2}
\label{514}
\ee
with
\be
Y^{N=2}=
\sum_{\a}{\left\vert G_{\alpha}\right\vert \over \vert G\vert} \,
Y\left(T_{\a},U_{\a},\bT_{\a},\bU_{\a}\right) \, ,
\label{515}
\ee
where $Y\left(T_{\a},U_{\a},\bT_{\a},\bU_{\a}\right)$
is given by eq. (\ref{71}). In eqs. (\ref{514}) and (\ref{515}),
$T_{\a}$ and $U_{\a}$ are the moduli corresponding to the
complex plane whose little group is $G_{\a}$, and
$b_i^{\a}$ are the beta-function coefficients of the $N=2$ orbifold
${\bf T}^6/G_{\a}$, which allow us to express the  $N=2$
contributions to the beta-function coefficients as follows:
\be
b_i^{N=2}=
\sum_{\a}b_i^{\a}\,  {\left\vert G_{\alpha}\right\vert \over \vert
G\vert}
\, .
\label{516}
\ee

A few remarks are in order here.
At the level of the ${\bf T}^6/G_{\a}$ orbifold, the moduli
$T_{\a}$ and $U_{\a}$ are completely free. When brought down to
the $N=1$ model, however, some of them might be constrained
in order to fit the discrete symmetry that is modded out. Therefore,
some of the arguments in  expressions (\ref{514}) and (\ref{515}) are
in general frozen to some specific values (more details about that
can be found in \cite{dkl}).

As far as stringy $N=1$ contributions are concerned,
no moduli dependence appears in the thresholds since none of the
corresponding twists
acts trivially on any plane inside ${\bf T}^6$. On the other hand, no
systematic factorization of a real function such as $\Gamma_{2,2}$ is
possible in the integrand of the first term of (\ref{511}).
Consequently, one cannot advocate holomorphicity and
analytic properties to determine the generic structure for this
threshold, as was done in section 3
(see eqs. (\ref{56}) and (\ref{61})). Nevertheless, in order to get a
better appreciation of its influence on the low-energy physics, we
can proceed to a numerical estimation in the
case of two models, namely the symmetric $Z_3$ and $Z_4$ orbifolds.
Our starting point is eq. (\ref{511}), that we can
recast in the following form:
\be
\Delta_{i}^{N=1}=
b_{i}^{N=1}\log {2\,e^{1-\gamma}\over \pi\sqrt{27}}+
\delta_i^{\vphantom N}\, ,
\label{517}
\ee
where
\be
\delta_{i}^{\phantom N}=
\int_{\cal F}{d^2\t\over \im}
\left(
d_i^{\phantom N}
(\t,\bar \t)-b_{i}^{N=1}
\right)-{k_i\over 4\pi}
\int_{\cal F}{d^2\t\over (\im)^2}
\, y(\t,\bar \t)
\label{517b}
\ee
and
$d_i(\t,\bar \t)$ and $y(\t,\bar \t)$ are model-dependent functions.

\vskip 0.3cm
{\sl (\romannumeral1) The symmetric $Z_3$ orbifold}

In this model,
the gauge group is $E_8\times E_6\times SU(3)$ and
the absence of $N=2$ sectors implies that
$\Delta^{N=2}_i =b^{N=2}_i = 0$. The beta-function coefficients
are ($b_i^{\vphantom{N=1}}\equiv b_i^{N=1}$)
$b_{E_8}=-90$, $b_{E_6}=72$ and $b_{SU(3)}=72$ respectively,
while the functions introduced here above read
 ($ q = e^{2 \pi i  \tau}$):
\begin{eqnarray}
d_{E_8} &=&-90  -
540\, q +
26460\,  \qb +
918540\,  q^{1\over3}\, \qb^{4\over3}
+ 158760 \, q\, \qb+
2963520\, \qb^2 +
{\cal O}\left(q^{r>2}\right)\, ,\cr
d_{E_6} &=& 72  +
2916\,  q^{1\over3}\, \qb^{1\over3} +
432\,  q +
27432\,  \qb +
2916\,  q^{4\over3} \, \qb^{1\over3} +
921456\,  q^{1\over3}\, \qb^{4\over3}\cr
&&+
164592 \, q\, \qb+
2963520\,  \qb^2 +
{\cal O}\left(q^{r>2}\right)\, ,\cr
d_{SU(3)} &=&d_{E_6}\, ,\cr
y&=& 108 -
\frac{3}{2}\, \frac{1}{\qb}-
9\, {q\over \qb} +
 15309\,  q^{1\over3}\, \qb^{1\over3} +
648\,  q +
150174 \, \qb +
 15309 \, q^{4\over3}\, \qb^{1\over3}\cr
&&+ 4496472\,  q^{1\over 3}\, \qb^{4\over3}
-9\, \frac{q^3}{\qb} + 9010444 \, q\,  \qb + 12070176\,  \qb^2
+ {\cal O}
\left(q^{r>2}\right)\, .
\end{eqnarray}
After numerical integration, taking into account
${\cal O}\left(q^{7\over 2}\right)$ terms,  we obtain:
\be
\delta_{E_8} \approx  -16.0  \; , \; \;
\delta_{E_6} =\delta_{SU(3)}\approx -5.0 \, .
\ee
Note that $\delta_{SU(3)}-\delta_{E_8} \approx 11.0$, in agreement
with \cite{kaplu}. In this case , due to the fact that
$d_{E_6} = d_{SU(3)}$ we can express the thresholds as for the $N=2$
models (eq. (\ref{66a})):
\begin{equation}
\Delta_i^{N=1} = b_i^{N=1} \, \Delta^{N=1}_{\vphantom i} -
Y^{N=1}_{\vphantom i}\, ,
\label{dec}
\end{equation}
where
\be
\Delta^{N=1}=
\log {2\,e^{1-\gamma}\over \pi\sqrt{27}}
+\delta\, .
\ee
Our numerical results lead to
$\delta\approx 0.07$ and $Y^{N=1}\approx 9.82$. However, this is not
to be
considered as a generic feature of $N=1$ sectors, and we will meet a
counter example bellow.
In the case under consideration, using eqs. (\ref{49}) and
(\ref{dec})
as well as the non-renormalization theorem \cite{kkpr} for
(\ref{three}), we can define
a common unification scale for all gauge couplings:
\be
M_{U} =M_P \,g_U  \,
\sqrt{\frac{2\,  e^{1-\gamma+\delta}} {\pi\sqrt{27}}}\,
\frac{1}{\sqrt{1+ \frac{Y^{N=1}}{16\, \pi^2} \, g_U^2}}\, ,
\label{UN1}
\ee
where
we introduced,
as previously, the effective field theory parameters $M_P$ and
$g_U= \frac{1}{g_i(M_U)}$ which  is the common coupling at the
unification scale. This coupling is again the one that was
introduced in section 2 as the ``renormalized" string coupling,
$g_{\rm renorm}$ (see eqs. (\ref{53}) and (\ref{55})).
Contrary to the $N=2$ case, here it is not possible to shift $M_U$ by
moving
the moduli.
Using our numerical results, we obtain:
\be
M_U \approx 5.4\times 10^{17}\times
g_U\times\frac{1}{\sqrt{1+0.06\times  g_U^2}}\,
{\rm GeV.}
\ee
 Thus the universal contributions lead to a small decrease of
the unification scale, which is of the order of $3\%$ for $g_U\sim
1$.

\vskip 0.3cm
{\sl (\romannumeral2) The symmetric $Z_4$ orbifold}

The gauge group is  now $E_8\times E_6\times SU(2)\times U(1)$.
Here the set of $N=2$ sectors is that of the symmetric $Z_2$
orbifold
and the corresponding thresholds are given by eqs.
(\ref{71}), (\ref{514}) and (\ref{515}).
On the other hand, beta-function coefficients read:
$b_{E_8}^{N=1}=-60$,
$b_{E_6}^{N=1}=36$, $b_{SU(2)}^{N=1}=12$, $b_{U(1)}^{N=1}=72$ and
$b_{E_8}^{N=2}=-30$,
$b_{E_6}^{N=2}=b_{SU(2)}^{N=2}=b_{U(1)}^{N=2}=42$, and
\ba
d_{E_8} &=&
-60 + 960 \, q^{1\over 2}\, \qb ^{1\over 2} - 240\,  q + 14040 \, \qb
+
 245760 \, q^{1\over4}\, \qb^{5\over4} +
{\cal O}\left(q^{2}\right)\, ,
\cr
d_{E_6} &=&
36 +768\, q^{1\over4}\, \qb^{1\over4}+ 144 \, q +  14424 \, \qb +
2496\,  q^{1\over2}\,  \qb ^{1\over2} +
1536 \, q^{5\over4}\, \qb^{1\over4}
\cr
&&+247296\,  q^{1\over 4}\,  \qb^{5\over4}+
{\cal O}\left(q^{2}\right)\, ,
\cr
d_{SU(2)} &=&
12 - 8\, \frac{q^{1\over 2}}{\qb^{1\over 2}} + 1024 \, q^\frac{1}{4}
\, \qb^\frac{1}{4}
+ 48 \, q + 8056 \, \qb + 2048 \, q^\frac{5}{4}\, \qb^\frac{1}{4} +
313344 \,  q^\frac{1}{4}\, \qb^\frac{5}{4} +
{\cal O}\left(q^{2}\right)\, ,
\cr
d_{U(1)} &=&
72 + 12\, \frac{q^{1\over 2}}{\qb^{1\over 2}} + 384 \,
q^\frac{1}{4}\,  \qb^\frac{1}{4}
+ 6240 \, q^{1\over 2}\, \qb^{1\over 2}
+ 288 \, q + 23976 \, \qb + 768 \, q^\frac{5}{4}\, \qb^\frac{1}{4}
\cr
&&+  148224 \,  q^\frac{1}{4}\, \qb^\frac{5}{4} +
{\cal O}\left(q^{2}\right)\, ,
\cr
y&=&12 -{1\over q} - 4\, \frac{q}{\qb}  +16 \, {q^{1\over
2}\over\qb^{1\over 2}}
+ 4096 \, q^{1\over4}\, \qb^{1\over4} + 48\,  q + 161128 \, q^{1\over
2}\, \qb^{1\over 2} + 81856 \, \qb
\cr
&&-4 \, \frac{q^2}{\qb}
 + 1220608 \, q^{1\over4}\, \qb^{5\over4} +
{\cal O}\left(q^{2}\right)
\, .
\ea
Again, numerical evaluation performed with the same accuracy as
before leads to:
\be
\delta_{E_8} \approx   -6.6 \; , \; \;
\delta_{E_6} \approx   4.0 \; , \; \;
\delta_{SU(2)}\approx  6.9  \; , \; \;
\delta_{U(1)}\approx   0.4 \, .
\ee
In this case the decomposition (\ref{dec}), which was usually adopted
in the literature
\cite{nilles}, does not hold for the $N=1$
contributions.

Putting together eqs. (\ref{511}),
(\ref{514}), (\ref{515}) and (\ref{517}), we obtain for the threshold
corrections of the $Z_4$ orbifold:
\be
\Delta_{i}^{\vphantom N}=
b_{i}^{N=1}\log {2\,e^{1-\gamma}\over \pi\sqrt{27}}+
\delta_i^{\vphantom N}
+b_{i}^{N=2}\,
\Delta\left(T_{3},U_{3},\bT_{3},\bU_{3}\right)- {1 \over 2}\,
Y\left(T_{3},U_{3},\bT_{3},\bU_{3}\right)
\, ,
\label{thz4}
\ee
where $\Delta$ and $Y$ are given by eqs. (\ref{67}) and (\ref{logd})
respectively. The decomposition (\ref{66a})
where $b_i$ are the full beta-function coefficients
is no longer valid,
and thus it is not possible to define a
unification scale common to all couplings.
In order to gain insight it is however interesting to determine the
scale
$M_U^{E_8 - E_6}$ where the $E_8$ and $E_6$ gauge couplings meet.
This scale can be found by following steps similar to those
introduced above.
It is expressed as a function of the moduli as well as of the common
value of the couplings at that scale:
$g_{E_8}^{\vphantom U}\left(M_U^{E_8 - E_6}\right)
=g_{E_6}^{\vphantom U}\left(M_U^{E_8 - E_6}\right)
=g^{E_8 - E_6}_U$. The latter is related to $\gs$
as usually, in a moduli-dependent way. Again, the minimum of this
unification scale is reached at $T_3=U_3=i$ with the result:
\begin{equation}
M_{U\ \scriptstyle\rm min}^{E_8 - E_6}
\approx 5.49\times 10^{17}\times g^{E_8 - E_6}_U\times
\frac{1}
{\sqrt{1 + 0.08 \times \left(g^{E_8 - E_6}_U\right)^2}}
\, {\rm GeV}\, .
\label{dz4min}
\end{equation}
Furthermore, eq. (\ref{49}) enables us to compute the splittings of
the
$SU(2)$ and $U(1)$ gauge couplings with respect to the $E_8$ and
$E_6$
ones, at that scale. We obtain:
\be
{1\over g^2_{SU(2)}\left(M_U^{E_8 - E_6}\right)}-
{1\over g^2_{E_8}\left(M_U^{E_8 - E_6}\right)}
\approx 0.035
\label{spl1}
\ee
and
\be
{1\over g^2_{E_8}\left(M_U^{E_8 - E_6}\right)}-
{1\over g^2_{U(1)}\left(M_U^{E_8 - E_6}\right)}
\approx 0.048
\, ,
\label{spl2}
\ee
which show that the relative splittings are of the order of
$4$ to $7 \%$.

\vskip 0.3cm
\setcounter{section}{5}
\setcounter{equation}{0}
\section*{\normalsize{\bf 5. Conclusions}}

\noindent
Let us now summarize our results. By using a method introduced in
\cite{kkb} that allows us to handle the infra-red problems,
we have determined the complete one-loop gauge coupling corrections
(eq. (\ref{51})) for general
heterotic four-dimensional models with at least $N=1$ supersymmetry.
These corrections contain both universal and group-factor dependent
terms. Our results for the latter are
in agreement with those obtained previously following a different
procedure
\cite{kaplu, dkl}, when evaluated within the same ultraviolet
renormalization scheme, here the
$\overline{DR}$ scheme. This shows that the relation between the
running gauge couplings of the low-energy field theory and the
string coupling does not depend
on the infra-red regularization prescription. It amounts to
the decoupling of the (infinite tower of) massive states and
allows for an unambiguous definition of string effective theory.
Such a conclusion could not have been drown without using
a consistent infra-red regulator.
Although our result has been established in the framework of an
infra-red regulator induced by a particular four-dimensional
curved background, we would have reached the same conclusions
within any other background possessing similar properties,
such as those listed in \cite{worm}.

Going beyond what has been achieved in previous studies
\cite{pr}, we have determined the moduli-dependent
universal part $Y\left(T,U,\overline{T},\overline{U}\right)$ of
the thresholds for the
class of $N=2$ four-dimensional theories that come from
torus compactification of six-dimensional $N=1$ ground states.
We have
obtained an explicit formula for these thresholds (eq. (\ref{68}))
which, thanks to the relation between gauge and $R^2$-term
renormalizations,
turns out to be
related to the quantity $N_{H}-N_{V}$.
This is {\it fully
determined}
as a consequence of {\it the anomaly cancellation} (gauge,
gravitational
and mixed) in the underlying six-dimensional theory. Therefore,
the whole class of models under consideration have
equal universal thresholds. This implies in particular that the
results of
\cite{hm} are actually more general, and provide the prepotential for
all
$N=2$ ground states that are toroidal compactifications of
six-dimensional $N=1$ vacua.

Using the method of orbits of the modular group, we have recasted the
integral representation of the above thresholds (\ref{71}) as a power
series expansion (eq. (\ref{Bexp})). This allowed us to analyse the
singularity behaviour of
$Y\left(T,U,\overline{T},\overline{U}\right)$: although this function
is continuous inside the moduli space (in contrast to what was
believed), its Laplacian, which is the one-loop correction to the
K\"ahler metric for moduli fields, diverges logarithmically around
enhanced-symmetry lines, but remains free of $\delta$-function
singularities. Finally, we have used the series representation for
analysing the asymptotics of the $N=2$ universal thresholds
(eqs. (\ref{BR}) and (\ref{BRR})). The leading behaviour is linear
with respect to each radius. This blow up as well as the bad
behaviour of the group-dependent contributions lead to the well known
decompactification problem in string theory.
This problem is cured by considering a class of
$N=4$ four-dimensional models, in which two supersymmetries are
spontaneously broken \cite{deco}.
Such models can be thought of as freely acting orbifolds where a
translation is performed in the $N=2$ invariant plane. Their
behaviour at large radii is drastically different from the standard
orbifolds. Indeed, the $N=4$ supersymmetry is restored in the
decompactification limit. Due to this restoration of the full
supersymmetry, the linear divergence of couplings with respect to the
radii is absent and this provides a solution to the
decompactification problem. It is finally interesting to observe that
models where the $N=4$ supersymmetry is spontaneously broken down to
$N=2$ and $N=1$ do also exist \cite{kka}. For these, string-string
dualities \cite{dua} and $D$-brane technology \cite{D} allow for a
better
understanding of the mass spectrum and the multiplicities of the
perturbative as well as non-perturbative BPS states \cite{vvd}.
This knowledge might eventually help for deriving the structure of
the
non-perturbative threshold corrections in physically interesting
$N=2$ and $N=1$ ground states \cite{kka}.

The results we obtained for the full threshold corrections enabled us
to analyse systematically the unification properties of various
models.
Concerning the $N=2$ ground states that we have studied in section 3,
we observed that the presence of the universal thresholds $Y$ (eq.
(\ref{logd})) leads to a lowering of the natural unification scale
with minima reached at the self-dual points $T=U=i,\rho$, and given
in
(\ref{dmumin}). More precisely, for $g_U^2=\frac{1}{2}$ we have a
$5\%$ decrease while for $g_U^2=1$ we can reach $10\%$, with
respect
to the case where these corrections are not taken into account.

The case of $N=1$
models is phenomenologically more interesting.
In our study we analysed $N=1$ orbifold constructions. In this case,
the above
achievements can be used in order to determine analytically the
($N=2$)-sector contributions to the thresholds with the results
(\ref{514}) and (\ref{515}).
Concerning the contributions originated from the $N=1$ sectors, no
general formula is available for the moment,
except for the Green--Schwarz term, which appears naturally in the
$S$-frame but does not play any role in string unification,
as we showed in section 2 by using the results of appendix A.
We therefore restricted our attention to the particular cases
of $Z_3$ and $Z_4$ symmetric orbifolds, which allowed us to draw some
interesting conclusions. For the $Z_3$ orbifold, the decomposition
(\ref{dec}) makes it possible to define a unification scale for all
couplings (eq. (\ref{UN1})), similar to the one that we introduced
here above
in the case of $N=2$
models
(eq. (\ref{dmu})), but moduli-independent.
Again the presence of universal thresholds reduces this
scale by a few percent. However, this decomposition
is  accidental and
does not apply in more general situations, as e.g. the $Z_4$
orbifold,
in contrast to some
general wisdom. Thus one cannot any longer define a common
unification scale for all couplings. It is however possible to
introduce a scale
where a pair of couplings meet. In the case of the $Z_4$ orbifold, we
determined that scale for the $E_8$ and $E_6$ couplings, and observed
that the relative splittings of the others were of the order of $4$
to
$7 \%$. This situation has to be compared to what happens in ordinary
grand unified theories where, it is always possible to chose a
scheme, namely the $\overline{DR}$ scheme, such that all couplings
are
unified at some scale \cite{IKT}.

Despite the various effects and contributions that one can advance in
order to
reduce the string unification scale, we should be aware that in the
models considered, this
scale
generally concerns groups that have little to do with
phenomenology, and that one has somehow
to break one of these groups, say
$E_6$, down to some subgroup, eventually leading to the
standard model.
In order to describe such a realistic situation in the framework of
strings,
it seems difficult to avoid the introduction of Wilson
lines \cite{KPR}.
Those will enhance the moduli space and allow for a better
exploration of the various symmetry-breaking possibilities.

\newpage
\vskip 0.3cm
\centerline{\bf Acknowledgements}

\noindent
C. Kounnas was  supported in part by EEC contracts
SC1$^*$-0394C and SC1$^*$-CT92-0789.
J. Rizos would like to thank the CERN Theory Division  for
hospitality
and  acknowledges financial
support from the EEC contract {ERBCHBGCT940634}.

\vskip 0.3cm
\setcounter{section}{0}
\setcounter{equation}{0}
\renewcommand{\theequation}{A.\arabic{equation}}
\section*{
\centerline{\normalsize{\bf Appendix A: The Green--Schwarz term in
$N=1$
string ground states}}}

We present here a discussion on the appearance of the Green--Schwarz
term
in $N=1$ string ground states \cite{dfkz} when string theory results
in the dilaton frame are matched with effective supergravity
calculations in the $S$-frame.

We will start with the tree-level plus one-loop bosonic action of the
heterotic
string in a generic $N=1$ ground state,
$S=S_{\rm tree}+S_{\rm one \ loop}$, where (we set $\a '=1$)
\be
S_{\rm tree}=
\int d^4 x\, \sqrt{G}\, e^{-2\Phi}
\left({1\over 2}\left( R+4(\partial \Phi)^2
-{1\over 12}H^2\right)-{k\over 4}F^2-
K^{\, (0)}_{T_\alpha \overline{T}_\beta}
\partial
T_\alpha \partial
\bT_\beta+ \cdots\right)\, ,
\label{A1}
\ee
\be
S_{\rm one \ loop}=
\int d^4 x\, \sqrt{G}
\left(-K^{\, (1)}_{T_\alpha \overline{T}_\beta}
\partial
T_\alpha \partial
\bT_\beta
-{k Z_F\over 4}  F^2+{Z_{F\widetilde{F}}\over 4} F
\widetilde{F}+{1\over 2}H^{\mu}X_{\mu}+\cdots\right)
\, .\label{A2}
\ee
Here we included a single gauge field, the moduli and the universal
sector.
As usual
\be
H^{\sigma}={1\over 3!}{\epsilon^{\mu\nu\rho\sigma}\over
\sqrt{G}}H_{\mu\nu\rho}\ , \ \ \tilde F^{\mu\nu}={1\over 2!}
{\epsilon^{\mu\nu\rho\sigma}\over \sqrt{G}}F_{\rho\sigma}\, ,
\label{A3}
\ee
and $X_{\mu}$ is a vector that depends on the moduli.
The coupling of the antisymmetric tensor to the moduli that arises at
one loop
is a direct descendant of the anomaly-cancelling Green--Schwarz term
in ten dimensions.
Going to the Einstein frame where $g_{\mu\nu}=e^{-2\Phi}G_{\mu\nu}$,
and
introducing the axion by\footnote{Note that the $X_{\sigma}$ term is
the one-loop correction to the tree-level definition of the axion.}
\be
e^{-4\Phi}H^{\mu\nu\rho}={\epsilon^{\mu\nu\rho\sigma}\over
\sqrt{g}}
\left(\pp_{\sigma} A+X_{\sigma}\right)\, ,
\label{A4}\ee
we obtain the dual action:
\ba
{\widetilde S}_{\, {\rm tree \ \& \ one \ loop}}=
\int d^4 x\, \sqrt{g}\,
\Bigg({1 \over 2}
R-(\pp\Phi)^2-\left(K^{\, (0)}_{T_\alpha \overline{T}_\beta}
+e^{2\Phi}K^{\, (1)}_{T_\alpha \overline{T}_\beta}\right)
\partial
T_\alpha \partial
\bT_\beta\cr
-{1\over 4}e^{4\Phi}(\pp
A+X)^2
-{k\over 4}\left(e^{-2\Phi}+Z_F\right)F^2
+{1\over 4}\left(k A+Z_{F\widetilde{F}}\right)F\widetilde{F}\Bigg)\,
{}.
\label{A5}
\ea

At tree level the $S$ field is simply $S=A+ie^{-2\Phi}$ and the
tree-level
K\"ahler potential is
$K_{\rm tree}=
-\log \Im S + K^{(0)}\big(
T_\alpha, \overline{T}_\beta
\big)$.
At one loop the $S$ field mixes with the moduli and is determined by
a
K\"ahler
potential of the form \cite{dfkz}
\be
K_{\, {\rm tree \ \& \ one \ loop}}=
-\log\Big(
\Im S+V\big(
T_\alpha, \overline{T}_\beta
\big)\Big)+K^{(0)}\big(
T_\alpha, \overline{T}_\beta
\big)\, ,
\label{A6}
\ee
where $K^{(0)}\big(
T_\alpha, \overline{T}_\beta
\big)=-\log \Im T_\a - \cdots$.
Compatibility of (\ref{A6}) with (\ref{A5}) implies that
\be
S=A+i\left(e^{-2\Phi}-V\right)
\ ,\ \ X_{\mu}=i\left(V^{\vphantom{(1)}}_{T_\alpha}\partial_{\mu}
T_\alpha-V^{\vphantom{(1)}}_{\overline{T}_\beta}
\pp_{\mu}\overline{T}_\beta\right)
\ ,\ \ K^{\, (1)}_{T_\alpha \overline{T}_\beta}=
-V^{\vphantom{(1)}}_{T_\alpha \overline{T}_\beta}\, .
\label{A7}\ee

Let us consider now the physical gauge coupling constant.
{}From string theory we can calculate the one-loop corrected
$i$th group factor coupling to be, in the dilaton frame,
(see eqs. (\ref{KKB}) and (\ref{51}))
\ba
{1\over g_{i,\,{\rm eff} }^2}&=&k_i\, e^{-2\langle
\Phi\rangle}+{b_i\over 16\, \pi^2}\log{M_s^2\over \mu^2}
+{\Delta_i\over 16\, \pi^2}
+ {b_i\over 16\, \pi^2}(2+2\g)\cr
&=&
k_i\left(\Im S+V\right)+{b_i\over 16\, \pi^2}\log{M_s^2\over
\mu^2}+{\Delta_i\over 16\, \pi^2}
+ {b_i\over 16\, \pi^2}(2+2\g)\, ,
\label{A9}
\ea
where $\mu$ is the infra-red scale, $\Delta_i$ are the thresholds
computed in
section 2 (eq. (\ref{51})) and $b_i$ the full beta-function
coefficients of the $N=1$ model: $b^{\vphantom r}_i=\sum_r
n^r_i \,  T^{\vphantom r}_i(r)-3 T(G^{\vphantom r}_i)$.

In the effective supergravity theory, at tree level, the gauge
couplings are given by the imaginary part of a holomorphic function
$f_i$, which in this case is $f_i=k_i \, S$. At one loop we have
\cite{dfkz, kl}
\ba
{1\over g_{i,\,{\rm eff} }^2}&=&k_i \, \Im S +{b_i\over 16\,
\pi^2}\left(
\log{\Lambda^2\over \mu^2}-\log \Im S \right)
+ {\rm \ constant}
\cr
&&+{1\over 16\, \pi^2}
\left(\Im f^{(1)}_i+c_i^{\vphantom{(0)}}\, K^{(0)}_{\vphantom{i}}
-2\sum_r
T_i^{\vphantom{(0)}}(r)\log \det Z^{(0)}_{r_i}\right)\, .
\label{A10}
\ea
Here $c^{\vphantom r}_i=\sum_r
n^r_i \,  T^{\vphantom r}_i(r)- T(G^{\vphantom r}_i)$
is the tree-level
(moduli-dependent) matrix
that multiplies the kinetic terms of matter in the representation $r$
of  $G_i$.
There is a scheme-dependent constant in (\ref{A10}), which for
the
$\overline{DR}$ scheme was computed in \cite{kaplu,pr},
and turns out to be
${b_i\over 16\, \pi^2}(2+2\g)$.
The calculation in the effective field theory was done with an
ultraviolet cut-off at $\Lambda$.
In \cite{kl} the cut-off was set to be the Planck scale $\Lambda
=M_P$,
which is the natural scale from the point of view of supergravity.
It was shown in \cite{kkpr} that in heterotic supersymmetric string
vacua the relation of the string scale to the Planck mass does not
receive corrections
in perturbation theory.
Thus
\be
M_P=M_s \, e^{-\langle \Phi\rangle}\, .
\label{A11}
\ee
However, if this relation is expressed in terms of $\Im S$ then it
does receive corrections already at one loop, since the relation of
$\Im S$
and $e^{-2\Phi}$ is modified order by order in perturbation theory.
Therefore, to first non-trivial order
\be
M_s^2\Big\vert_{\rm one \ loop}={M_{P}^2\over \Im S +V}\, .
\label{MP}
\ee
Choosing $\Lambda=M_P$ does not affect the couplings  at one loop.
Nevertheless, in order to profit from the non-renormalization theorem
we
can choose
the ultraviolet cut-off for the effective field theory to be defined
by
(\ref{MP}), which effectively resums all relevant higher-loop effects
due to the renormalization of the relation between $e^{-2\Phi}$ and
${\rm
Im}S$.
This is a slightly different scheme than the one
adopted in \cite{kl}. It is similar to loop computations in
QCD, where
the judicious choice of the relevant scale for a process resums some
higher-order effects.

Comparing now (\ref{A9}) and (\ref{A10}) we obtain
to next-to-leading order in
$\Im S$:
\be
\Im f^{(1)}_i=-c_i^{\vphantom{(0)}}\, K^{(0)}_{\vphantom{i}}
+2\sum_r
T_i^{\vphantom{(0)}}(r)\log \det Z^{(0)}_{r_i}
+16\, \pi^2 k_i^{\vphantom{(0)}} \,V + \Delta_i^{\vphantom{(0)}}\, .
\label{A12}
\ee
All terms in (\ref{A12}) are independent of the $S$ field.
Moreover \cite{kl} since $f^{(1)}$ is holomorphic due to $N=1$
supersymmetry,
the right-hand side has to be a holomorphic function.
Thus the non-holomorphicity of the tree-level kinetic terms has to
cancel
the one-loop non-holomorphicity of the couplings.
This is a reflection of the cancellation of the K\"ahler anomaly in
the effective field theory
\cite{dfkz}.

One more comment is in order here. One-loop calculations in the
effective
theory
reproduce derivatives of the function $V$. On the other hand, the $S$
field
is defined in terms of $V$ itself. A harmonic shift of $V\to V+h+\bar
h$, where
$h$ is holomorphic, can be absorbed in a holomorphic redefinition of
$S$.
Thus the coupling in the $S$ frame is defined up to holomorphic
redefinitions.
This is important in the context of the threshold corrections
generated from
$N=1$ sectors.

Finally, formula (\ref{A10}) refers to generic heterotic superstring
vacua. If
one restricts to orbifold compactifications, and use the extra
information
they provide for the K\"ahler potential \cite{dfkz, kl}, $V$ can be
determined (up to holomorphic redefinitions):
\be
V =
\frac{1}{16\, \pi^2} \left(\Delta^{GS}
 +
Y
\right)
\ee
with
\be
\Delta^{GS}_{\vphantom i}   =
\sum_{\alpha} \delta_\alpha^{GS}\left(
\log\Im T_\alpha + \log\Im U_\alpha\right)
\ee
the advertised Green--Schwarz term,
where the sum extends over the moduli that are not fixed by the
orbifold
action, and
$Y$ is the universal threshold correction appearing in
(\ref{52}) and (\ref{51}). Both $Y$ and  $\Delta^{GS}$
are universal in that they do not depend on the gauge group factor;
however their origin is respectively {\it stringy} and {\it
field-theoretical},
and they should not be confused.
The numerical constants $\delta_\alpha^{GS}$ can be
calculated
from the spectrum of the model. An explicit calculation of
these quantities
for various orbifold models was presented in  \cite{dfkz, kl}.
For $N=2$ models we have  $\delta^\a_{GS}=0$ and thus
\be
V^{N=2}= {1\over 16\, \pi^2}  Y\left(T,U,\bT,\bU\right)
\label{A15}
\ee
with
$Y\left(T,U,\bT,\bU\right)$
given in (\ref{71}).
In situations with $N=1$ supersymmetry, $\delta^\gamma_{GS}$
can be evaluated  for specific models \cite{kl}.
They turn out to vanish for $Z_2\times Z_2$ orbifolds, while for the
symmetric $Z_3$ orbifold we have
$\delta^1_{GS} = \delta^2_{GS} =\delta^3_{GS} =30$:
\be
\Delta^{GS} = 30 \sum_{\alpha=1}^3 \log \Im T_\alpha\, .
\label{ey}
\ee
In the case of the symmetric $Z_4$ orbifold, the result is
$\delta^3_{GS}=0$
for the $N=2$ plane that is left unrotated by the orbifold twist, and
$\delta^1_{GS} = \delta^2_{GS} =30$ for the other two planes. Thus
\be
\Delta^{GS} = 30 \sum_{\alpha=1}^2\log \Im T_\alpha\, .
\ee

Our last comment concerns the issue of unification. In section 2,
working in the dilaton frame, we
introduced
a renormalized coupling which plays the role of unification
coupling, when unification exists. This coupling is related to
$\gs = \exp \langle \Phi \rangle$ as shown in eq. (\ref{55}). Moduli
dependence enters through $Y$ and propagates to the unification scale
when $M_s$
is expressed in terms of $M_P$ (see eq. (\ref{A16})). In the
$S$-frame, the same renormalized coupling can be introduced, however
expressed in terms of
$\Im S$:
\ba
{1\over g_{\rm renorm}^2}\Bigg\vert_{\rm one \ loop} &=& \Im S
+V -{Y \over 16\, \pi^2}\cr
&=&  \Im S +{\Delta^{GS} \over 16\, \pi^2}\, .
\label{grenS}
\ea
All moduli dependence is now contained in $\Delta^{GS}$. However, if
one uses eq.
(\ref{grenS}) to recast (\ref{MP}) in terms of $g_{\rm renorm}$, we
obtain (\ref{A16})
as in the dilaton frame, and the moduli dependence appears again
through $Y$. Therefore, eq. (\ref{A16}) do not depend on the frame,
and so are the conclusions about the unification scale, which turns
out to be affected
by $Y$ but {\it is not sensitive to $\Delta^{GS}$}.
\vskip 0.3cm
\setcounter{section}{0}
\setcounter{equation}{0}
\renewcommand{\theequation}{B.\arabic{equation}}
\section*{\normalsize{
\centerline{\bf Appendix B: Explicit formulas for the universal
thresholds and asymptotics}}}

The purpose of this appendix is to evaluate, in terms of a multiple
series expansion, the universal thresholds for $N=2$ models that are
toroidal compactifications of $N=1$ six-dimensional heterotic vacua,
and analyse their asymptotic behaviours. Our starting point is eq.
(\ref{logd}). By using (\ref{66}), we can recast expression (\ref
{logd}) as follows:
\ba
Y\left(T,U,\bT,\bU\right)&=&{1\over 12}\int_{\cal F}{d^2\t\over
\im}\,
\Bigg(\Gamma_{2,2}\left(T,U,\bT,\bU\right)
\left(\bE_2-{3\over \pi\im}\right){\bE_4\, \bE_6\over \bar
\eta^
{24}}+264
\Bigg)\cr
&&-22\left(
\log\left(\big|\eta(T)\big|^4 \big|\eta(U)\big|^4
\Im T  \Im U\right)+\log{8\pi \, e^{1-\gamma}\over \sqrt{27}}
\right)
\cr
&&+{1\over 3}
\log\big\vert j(T)-j(U)\big\vert
\, .
\label{Blogd}
\ea
The remaining integral in (\ref{Blogd}) can be evaluated using
the results of \cite{hm} with \footnote{Here we use the convention
$\Theta(0) =\frac{1}{2}$.}:
\ba
I\left(T,U,\bT,\bU\right)&=&{1\over 12}\int_{\cal F}{d^2\t\over
\im}\,
\Bigg(\Gamma_{2,2}\left(T,U,\bT,\bU\right)
\left(\bE_2-{3\over \pi\im}\right){\bE_4\, \bE_6\over \bar
\eta^
{24}}+264
\Bigg)\cr
&=&{1 \over 3} \Re \Bigg(
-24 \sum_{k>0}
\left(
11 \, {\cal L}i_1 \left(e^{2\pi i kT}\right)
-{30 \over \pi \Im T \Im U}\,
{\cal P}(kT)
\right)\cr
&&\ \ \ \ \ \ \ \ \ \,  -\, 24 \sum_{\ell >0}
\left(
11 \, {\cal L}i_1 \left(e^{2\pi i \ell U}\right)
-{30 \over \pi \Im T \Im U}\,
{\cal P}(\ell U)
\right)\cr
&&\ \ \ \ \ \ \ \ \ \, + \sum_{k>0,\, \ell >0}
\left(
\tilde c (k\ell ) \, {\cal L}i_1 \left(e^{2\pi i (kT+\ell U)}\right)
-{3\, c(k\ell ) \over \pi \Im T \Im U}\,
{\cal P}(kT+\ell U)
\right)\cr
&&\ \ \ \ \ \ \ \ \ \, +
{\cal L}i_1 \left(e^{2\pi i\big(
\Re T - \Re U+i
\vert
\Im T - \Im U
\vert\big)}\right)\cr
&&\ \ \ \ \ \ \ \ \ \,
-{3 \over \pi \Im T \Im U}\,
{\cal P}\Big(
\Re T - \Re U+i
\vert
\Im T - \Im U
\vert\Big)\Bigg)\cr
&&+
{60\, \zeta(3) \over \pi^2 \Im T \Im U}+22\left(
\log(\Im T \Im U)
+\log{8\pi \, e^{1-\gamma}\over \sqrt{27}}\right)\cr
&&+ \left(
{4\pi \over 3}{(\Im U)^2 \over \Im T}-{22 \, \pi \over 3}
\Im U- 4\pi \Im T
\right)\Theta (\Im T - \Im U)
\cr
&&+ \left(
{4\pi \over 3}{(\Im T)^2 \over \Im U}-{22 \, \pi \over 3}
\Im T- 4\pi \Im U
\right)\Theta (\Im U - \Im T)
\, .\label{BI}
\ea
Here $c(n)$ and $\tilde{c}(n)$ are the coefficients of the
Laurent expansions:
\be
\frac{E_4 \, E_6}{\eta^{24}} = \sum_{n=-1}^{\infty} c(n) \, q^n
= \frac{1}{q}-240  - 141444 \, q  - 8529280 \, q^2+\cdots
\ee
and
\be
\frac{E_2\, E_4 \, E_6}{\eta^{24}}=\sum_{n=-1}^{\infty}\tilde{c}(n)
\, q^n
= \frac{1}{q}-264  - 135756 \, q - 5117440\, q^2+\cdots\, .
\ee
The function ${\cal P}(x)$ is defined by
\begin{equation}
{\cal P}(x)= \Im x\, {\cal L}i_2 \left(e^{2\pi i x}\right)+
\frac{1}{2\pi}\, {\cal L}i_3\left(e^{2\pi i x}\right)\, ,
\end{equation}
and ${\cal L}i_j$ are the polylogarithms
\begin{eqnarray}
{\cal L}i_1(x) & = & \sum_{j=1}^{\infty}\frac{x^j}{j} = - \log(1-x)\,
,\\
{\cal L}i_2(x) & = & \sum_{j=1}^{\infty}\frac{x^j}{j^2} \, ,\\
{\cal L}i_3(x) & = & \sum_{j=1}^{\infty}\frac{x^j}{j^3} \, .
\end{eqnarray}
The above integral $I$ is logarithmically divergent when $T\to U$;
the singularity arises from the term
$
{1 \over 3} \Re {\cal L}i_1 \left(e^{2\pi i\big(
\Re T - \Re U+i
\vert
\Im T - \Im U
\vert\big)}\right)$. This very divergence cancels the one appearing
in
$Y$ through
$
{1\over 3}
\log\big\vert j(T)-j(U)\big\vert
$. Indeed we can use the product representation
\begin{equation}
j(T) - j(U) = e^{-2\pi i T} \prod_{k>0,\, \ell >-2}\left(1 -
e^{2\pi i(kT+\ell U)} \right)^{\hat{c}(k \ell )}
\label{jexp}\, ,\label{Bj}
\end{equation}
where $\hat{c}(n)$ are defined by\footnote{We use
the notation $j(\tau)$ or $j(q)$ with $q=e^{2\pi i \tau}$.}
\begin{equation}
j(q) -744 = \sum_{n=-1}^{\infty} \hat{c}(n) \, q^n =
\frac{1}{q} +196884 \, q + 21493760\,
q^2+\cdots \, .
\end{equation}
Putting everything together we obtain:
\ba
Y&=&\Re
\sum_{k>0,\, \ell >0}
\left(
{\tilde c (k\ell )-\hat c (k\ell )
\over 3}\, {\cal L}i_1 \left(e^{2\pi i (kT+\ell U)}\right)
-{c(k\ell ) \over \pi \Im T \Im U}\,
{\cal P}(kT+\ell U)
\right)
\cr &&+
{60\over \pi^2 \Im T \Im U}\, \Bigg(\zeta(3)+4\pi
\Re \sum_{k>0}{\cal P}(kT) +4\pi
\Re \sum_{\ell >0}{\cal P}(\ell U)
\Bigg)
\cr &&-
{1 \over \pi \Im T \Im U}\Re
{\cal P}\Big(
\Re T - \Re U+i
\vert
\Im T - \Im U
\vert\Big)\cr
&&+ \left(
{4\pi \over 3}{(\Im U)^2 \over \Im T}
+ 4\pi \Im T
\right)\Theta (\Im T - \Im U)
\cr
&&+ \left(
{4\pi \over 3}{(\Im T)^2 \over \Im U}
+ 4\pi \Im U
\right)\Theta (\Im U - \Im T)
\, .\label{Bexp}
\ea

The above expression enables us to check that
$Y$ is finite and continuous at $T=U$
for finite $T$ and $U$. Concerning
$\partial_{T}\, \partial_{\overline{T}}\, Y$,
it is obvious that potential singularities may arise from the term
$
{-1 \over \pi \Im T \Im U}\Re
{\cal P}\Big(
\Re T - \Re U+i
\vert
\Im T - \Im U
\vert\Big)
$
as well as from the $\Theta$-functions. The latter turn out to give
regular terms while the former leads to (\ref{dY}).

Finally, by using (\ref{Bexp}), various asymptotic behaviours can be
studied.
We restrict again to the case where $\Re T= \Re U=0$.
The limit $R_1= R_2=R \to \infty$
was derived in \cite{pr} with the result:
\be
Y(R) = 4 \pi R^2 + \frac{60  \, \kappa}{\pi R^2} +
{\cal O}\left(e^{-\pi R^2}\right)\, ,
\label{BR}
\ee
where
\be
\kappa = \frac{2}{\pi^2}\zeta(4)
+\sum_{j>0}\left(\frac{\coth\pi j}{\pi}
\frac{1}{j^3} + \frac{1}{\sinh^2\pi
j}\frac{1}{j^2}\right)\approx0.61\, .
\ee
Following similar steps one can derive the asymptotic expansions for
$\Im T, \, \Im U\to \infty$ with the ratio kept fixed. This amounts
to taking
$R_2\to\infty$, while $R_1 = \sqrt{\Im T / \Im U}$ is finite.
We obtain:
\be
Y(R_1,R_2)=\left\{ \begin{array}{l}
{4\pi \over 3}\,  R_2
\left({1\over R_1^3}+3\, R_1
\right)\Theta\big(R_1-1\big)+
{4\pi \over 3}\,  R_2
\left(R_1^3+ {3\over R_1}
\right)\Theta\big(1-R_1\big)\cr
+{60 \over \pi^2 R_2^2}\, \zeta(3)
+{\cal O}\left(e^{-\pi  R_2^2}\right)
{\rm \ \ if \ } R_1 \ne 1 \cr
\cr
{16\, \pi \over 3}\,  R_2
+{119 \over 2 \pi^2 R_2^2}\, \zeta(3)
+{\cal O}\left(e^{-\pi  R_2^2}\right)
{\rm \ \ if \ } R_1 = 1
\, ,\end{array}
\right.
\label{BRR}
\ee
from which it appears that $Y$ is not continuous at $T=U$ when the
radii become large. Similar results can be reached for $R_1 \to
\infty$, $R_2$ finite; then the discontinuity appears at $R_2=1$,
corresponding to $T={1\over U}$.

\begin{centering}
\begin{figure}[ht]
\epsfxsize=14.5cm
\epsfbox[40 170 560 620]{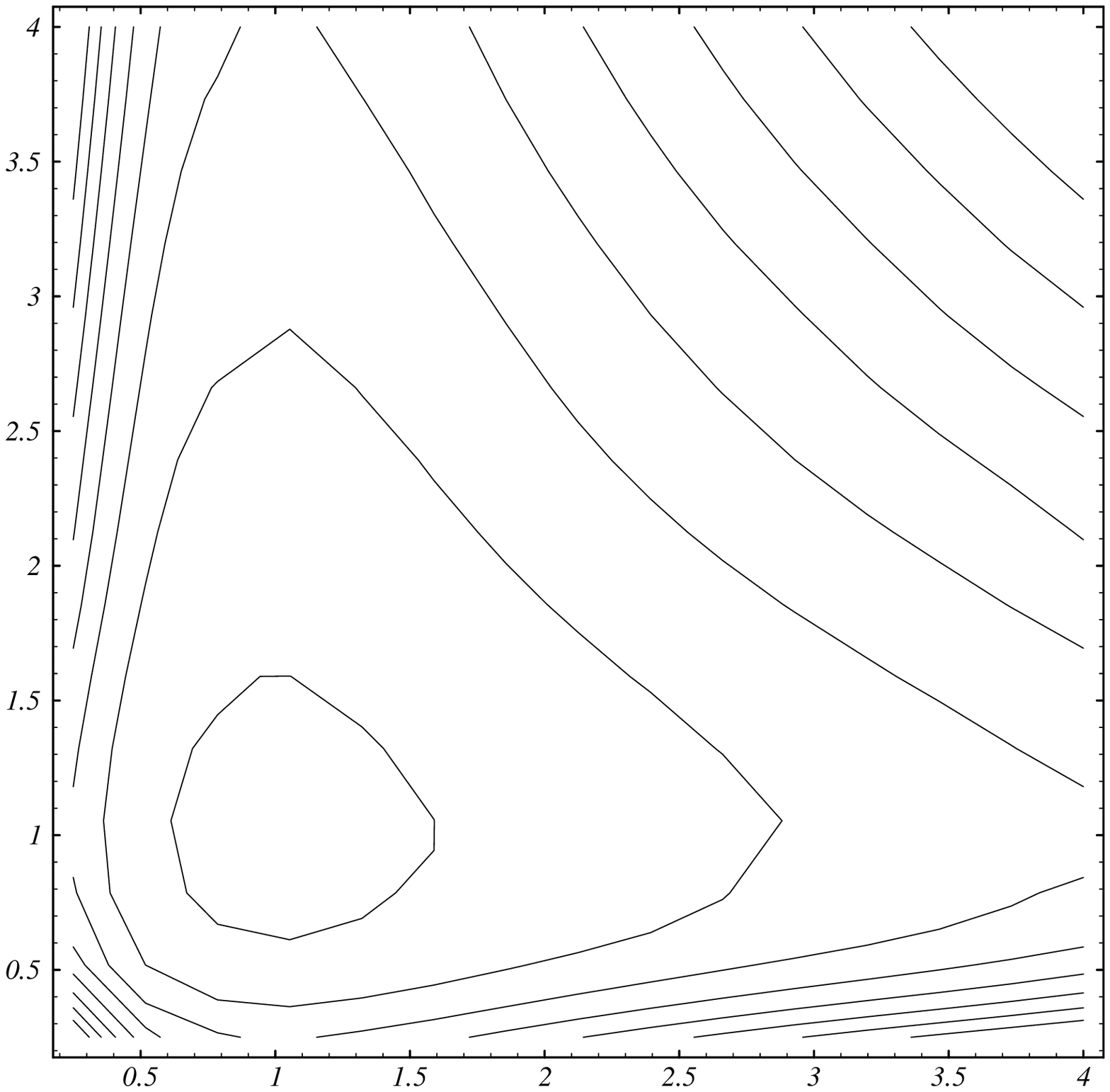}
\caption{ Contour plots of the universal thresholds $Y(R_1,R_2)$
as a function of the internal radii $R_1$ and $R_2$.}
\vskip -1.5cm\hskip11.cm$R_1$
\vskip-11.cm\hskip 0.5cm$R_2$
\vskip 6.5cm\hskip5.3cm $ $
\vskip -.7cm\hskip5.cm $30$
\vskip -2.8cm\hskip6.5cm $50$
\vskip -2.2cm\hskip7.3cm $70$
\vskip -2.0cm\hskip8.9cm $90$
\vskip -2.2cm\hskip10.7cm $130$
\vskip -2.0cm\hskip12.5cm $150$
\vskip 13.cm
\end{figure}
\end{centering}

We have also  calculated  $Y$ numerically. One can verify that the
$\Re T$, $\Re U$ dependence is very weak and leads to small
oscillations
of the order $<2\%$ around the $\Re T = \Re U =0$ values. The results
for
$\Re T = \Re U =0$ are presented in figure 1.
As expected, there is another minimum at the other self-dual point
$T=U=\rho$.
It has the same depth as the one at $T=U=i$ discussed above, to the
accuracy of its numerical evaluation. We come to the conclusion
that $Y(R_1,R_2)\ge Y(1,1)$ where $Y(1,1) \approx 24.4$ is the value
of
the minimum at the point $R_1=R_2 =1$.

\end{document}